\documentclass[12pt]{article}
\usepackage{graphicx}
\usepackage{epsfig}
\usepackage{epsf,amsfonts,amssymb}
\newcommand{\rom}[1]{\mathrm{#1}}

\newcommand{\beq}{\begin{equation}}
\newcommand{\eeq}{\end{equation}}
\newcommand{\beqa}{\begin{eqnarray}}
\newcommand{\eeqa}{\end{eqnarray}}
\newcommand{\beqar}{\begin{eqnarray*}}
\newcommand{\eeqar}{\end{eqnarray*}}

\newcommand{\labell}[1]{\label{#1}} 
\newcommand{\reef}[1]{(\ref{#1})}

\newcommand{\eg}{{\it e.g.,}\ }
\newcommand{\ie}{{\it i.e.,}\ }

\begin{document}
\setlength{\unitlength}{1mm}

\thispagestyle{empty}
\rightline{
\hfill \small hep-th/0412130}
\vspace*{2cm}

\begin{center}
{\bf \Large Non-supersymmetric black rings as thermally excited
supertubes}\\

\vspace*{1.4cm}

{\bf Henriette Elvang,}$^1\,$
{\bf Roberto Emparan,}$^{2,\,3}\,$
{\bf Pau Figueras}$^3\,$

\vspace*{0.3cm}

{\it $^1\,$Department of Physics, University of California, Santa
Barbara, CA 93106-9530, USA}\\[.3em]
{\it $^2\,$Instituci\'o Catalana de Recerca i Estudis Avan\c cats (ICREA)}\\[.3em]
{\it $^3\,$Departament de F{\'\i}sica Fonamental, and}\\
{\it C.E.R. en Astrof\'{\i}sica, F\'{\i}sica de Part\'{\i}cules i Cosmologia,}\\
{\it Universitat de Barcelona, Diagonal 647, E-08028 Barcelona, Spain}\\[.3em]

\vspace*{0.2cm}
{\tt elvang@physics.ucsb.edu, emparan@ub.edu,}\\
{\tt pfigueras@ffn.ub.es}

\vspace{.8cm} {\bf ABSTRACT}
\end{center}

We construct a seven-parameter family of supergravity solutions that
describe non-super\-symmetric black rings and black tubes with three
charges, three dipoles and two angular momenta. The black rings have
regular horizons and non-zero temperature. They are naturally
interpreted as the supergravity descriptions of thermally excited
configurations of supertubes, specifically of supertubes with two
charges and one dipole, and of supertubes with three charges and two
dipoles. In order to fully describe thermal excitations near
supersymmetry of the black supertubes with three charges
and three dipoles a more general family of black ring solutions is
required.

\noindent

\vfill \setcounter{page}{0} \setcounter{footnote}{0}
\newpage

\tableofcontents

\newpage

\setcounter{equation}{0}
\section{Introduction}

Black rings are a fascinating outcome of recent studies of
higher-dimensional gravity. They show that several classic results of
black hole theory cannot be generalized to five dimensions: black rings
have non-spherical horizon topology $S^1\times S^2$, and their mass and
spin are insufficient to fully distinguish between them and between
other black holes of spherical topology \cite{ER,RE}.

The remarkable progress in the string theory description of black holes
had not hinted at the existence of black rings. So, initially, black
rings appeared to be uncalled-for objects, and their role in string
theory was unclear. A step to improve the understanding of black rings
in string theory was taken in ref.~\cite{EE} (following \cite{HE}),
where a connection was found between black rings with two charges and
another class of objects of recent interest in string theory, the
so-called supertubes \cite{MT,EMT}. More recently this connection has
been significantly strengthened and extended with the discovery in
\cite{EEMR1} of a supersymmetric black ring of five-dimensional
supergravity, with a regular horizon of finite area. This has prompted
the further study of supersymmetric rings, including the
generalization to three-charge solutions \cite{EEMR2,BW,JGJG2} and other
extensions and applications \cite{JGJG1,BK2,TO,BWW,CGMS}. The authors of
\cite{CGMS} have actually succeeded in providing a statistical counting
of their Bekenstein-Hawking entropy. It is naturally interesting to try
to extend these results to include near-supersymmetric black rings. 

In this paper, we present a seven-parameter family of non-supersymmetric
black ring solutions which generalize the ones studied in
ref.~\cite{EE}. The new solutions describe black rings with three
conserved charges, three dipole charges, two unequal angular momenta,
and finite energy above the BPS bound. 
We are motivated by the wish to understand the microscopic nature of
the thermal excitations of two- and three-charge supertubes. We argue
that the near-supersymmetric limits of the black rings in this paper 
can be interpreted as thermally excited supertubes with two charges
and one dipole, or thermally excited supertubes with three charges and
two dipoles. 
As a further motivation, note that --- contrary to spherical black
holes --- the black rings carry non-conserved charges (the dipole
charges). As such the non-supersymmetric black ring solutions
provide an exciting laboratory for examining new features of black
holes, for instance the appearance of the non-conserved charges in the
first law of black hole thermodynamics \cite{RE,CH}. 

We find the non-supersymmetric black rings by solution-generating
techniques (boosts and U-dualities). This was also the approach in
\cite{EE,HE}, where the neutral five-dimensional black ring was first
uplifted to six dimensions, to become a black tube.\footnote{Throughout
this paper we refer to the same object as a ring (in the
five-dimensional description) or as a tube, when lifted to six or more
dimensions.} Then a sequence of solution-generating transformations
yielded new two-charge black tube solutions with the same charges as a
supertube \cite{MT,EMT}. In the supersymmetric limit the area of these
black tubes vanishes and one recovers the supergravity description of a
two-charge supertube \cite{EMT}. Thus these charged black rings can
be regarded as the result of thermally exciting a supertube. A
limitation of the charged black rings built in \cite{EE} is that
their supersymmetric limit could only yield supertubes with half the
maximum value of the angular momentum, instead of the whole range of
angular momenta that supertubes can have. This shortcoming is
automatically resolved in this paper. The additional parameters in our
new solutions allow us to construct thermal deformations for supertubes
with angular momenta covering precisely the entire physically permitted
range. 

The extra parameters in our solutions come from choosing a more general
seed solution to which the generating transformations are applied. While
in \cite{EE} the seed solution was the neutral black tube, we here use
the dipole black tubes of \cite{RE} as seed solutions. 

The additional degrees of freedom in the dipole solutions in fact allow
us to construct non-supersymmetric black rings with three charges and
three dipoles. Spherical five-dimensional black holes with three charges
are the most thoroughly studied black holes in string theory
\cite{review}, so having three-charge black rings should be instrumental
to develop the proposal in ref.~\cite{EE} for a microscopic
understanding of non-uniqueness and non-spherical topologies. A
different motivation to further study D1-D5-P configurations is Mathur's
programme to identify string microstates as non-singular horizonless
solutions \cite{mathur}. In fact this led the authors of \cite{BK,IB} to
independently conjecture the existence of supersymmetric black rings.

The supersymmetric limit of our solutions can only reproduce a
supersymmetric ring with three charges and at most two dipoles. The
complete three-charge/three-dipole black rings presented in
\cite{EEMR2,BW,JGJG2}, and the minimal supersymmetric ring of
\cite{EEMR1}, are not limits of the solutions in this paper. Indeed,
this becomes obvious by simply counting parameters. The supersymmetric
rings of \cite{EEMR2,BW,JGJG2} have seven independent parameters. This
is the same number as in the solutions in this paper, but in the latter,
one of the parameters measures the deviation away from supersymmetry. It
appears that in order to find the appropriately larger family of
non-supersymmetric black ring solutions one should start with a more
general seed, presumably a dipole black ring with two independent
angular momenta. Nevertheless, the solutions we present here seem to be
adequate to describe thermal excitations near supersymmetry of
supertubes with two charges and one dipole, and of some supertubes with
three charges and two dipoles.

The rest of the paper is organized as follows: in the next section we
discuss the general problem of how to construct the solutions, and then
present them and compute their main physical properties. In section
\ref{sec:extsusy} we analyze the extremal and supersymmetric limits of
these black rings. Section \ref{sec:minimal} analyzes the particular
case of solutions to minimal five-dimensional supergravity. In section
\ref{sec:d1d5p} we study black rings as thermally excited D1-D5-P
supertubes, and consider in particular the cases of tubes with two
charges and one dipole, and three charges and two dipoles. We also study
the decoupling limit. We conclude in section \ref{sec:discuss} with a
discussion of the consequences of our results. In appendices \ref{appA}
and \ref{appB} we provide the details for the form fields in the
solutions, and in appendices \ref{appC} and \ref{appD} we study the limits
where the solutions reproduce spherical rotating black holes (at zero
radius), and black strings (at infinite radius).


\setcounter{equation}{0}
\section{Non-supersymmetric black rings with three charges}
\label{sec:11dsoln}

In this section we first describe the sequence of boosts and dualities
that we exploit to generate the charged black ring solutions. The idea
is to follow the same path that yielded the three-charge rotating black
hole \cite{BMPV,BLMPSV,CY,tseytlin}, but this procedure becomes quite
more complex and subtle when applied to black rings. We then present the
solution in its most symmetric form, as an eleven dimensional
supergravity supertube with three M2 charges and three M5 dipole
charges, and analyze its structure and physical properties.

\subsection{Generating the solution}

We begin by reviewing the process followed in \cite{EE} to obtain
two-charge black rings, and why a problem arises when trying to add a
third charge. Starting from a five-dimensional neutral black ring that
rotates along the direction $\psi$, add to it a flat direction $z$ to
build a six-dimensional black tube. This is then embedded into
IIB supergravity by further adding four toroidal directions
$z_1,z_2,z_3,z_4$ (which will play little more than a spectator
role in the following). Now submit the solution to the following sequence of
transformations: IIB S-duality;
boost along $z$ with rapidity parameter $\alpha_1$; T-duality
along $z$; boost along $z$ with parameter $\alpha_2$; T-duality along
$z_1$; S-duality;
T-duality along $z_1,z_2,z_3,z_4$; S-duality; T-duality
along $z$; T-duality along $z_1$; S-duality.
Schematically,\footnote{The first and last S-dualities, and the two T(1)
dualities, are introduced simply to have both the initial and final
solutions later in this subsection as configurations of a D1-D5 system.}
\beq
S\to\mathrm{Boost}_{\alpha_1} (z) \to T(z) \to \mathrm{Boost}_{\alpha_2} (z)
\to T(1) \to S \to
T(1234) \to S \to T(z)\to T(1) \to S\,.
\label{chain}\eeq
As a result we obtain a non-supersymmetric solution of IIB supergravity.
The system has two net charges corresponding to D1 and D5-branes, and
there is a dipole charge from a Kaluza-Klein monopole (kkm). The branes
are arranged as
\beqa
\begin{array}{clccccccl}
\alpha_1  &\mbox{D5:} \,\, & 1  & 2  & 3  & 4  & z  & \_ & \, \\
\alpha_2  &\mbox{D1:} \,\, & \_ & \_ & \_ & \_ & z  & \_ & \, \\
(\alpha_1 \alpha_2)    &\mbox{kkm:}\,\, & 1  & 2  & 3  & 4  & (z) & \psi & \, .
\end{array}
\eeqa
Following ref.~\cite{EEMR2}, we use uppercase letters to denote brane
components with net conserved charges (D1, D5), and lowercase for dipole
brane charges (kkm). If the parameters on the left are set to zero then
the corresponding brane constituent disappears, \eg when either of the
$\alpha_i$ is zero the corresponding D-brane is absent, and the kkm
dipole, which is fibered in the $z$ direction, vanishes. Note that the
dipole is induced as a result of charging up the tube, and would not be
present if instead one charged up a spherical black hole. Ref.~\cite{EE}
argued that these black rings describe thermally excited supertubes:
configurations where D1 and D5 branes are `dissolved' in the worldvolume
of a tubular KK monopole.

It is natural in this context to try to have a third charge on the ring,
coming from momentum $P$ propagating along the tube direction $z$. In
order to endow the system
with this charge, one might try to perform a third boost on the
solution. However, as discussed in \cite{EE}, the kkm fibration along
the direction $z$ is incompatible with such a boost. A naive application
of a boost transformation to the solution results into a globally
ill-defined geometry with Dirac-Misner string singularities (a geometric
analogue of Dirac strings, to be discussed in detail later in this
section). 

To overcome this problem, in this paper we choose to start from a
different seed solution which already contains three dipole charges, with
parameters $\mu_i$, but no net conserved charges \cite{RE}. Beginning
now from a black tube with dipole charges d1, d5 and kkm given as
\beqa
\begin{array}{llccccccl}
\mu_1   &\mbox{kkm:}\,\, & 1  & 2  & 3  & 4  & (z) & \psi & \, \\
\mu_2   &\mbox{d5:} \,\, & 1  & 2  & 3  & 4  & \_ & \psi & \, \\
\mu_3   &\mbox{d1:}\,\, & \_ & \_ & \_ & \_ & \_ & \psi & \, ,
\end{array}
\label{seedarray}
\eeqa
and acting with the sequence \reef{chain} followed by a boost $\alpha_3$
along $z$, we obtain a black tube with the same dipole charges but now
also D1-D5-P net charges. The branes are arranged as 
\beqa
\begin{array}{clccccccl}
\alpha_1  &\mbox{D5:} \,\, & 1  & 2  & 3  & 4  & z  & \_ & \, \\
\alpha_2  &\mbox{D1:} \,\, & \_ & \_ & \_ & \_ & z  & \_ & \, \\
\alpha_3  &\mbox{P:}  \,\, & \_ & \_ & \_ & \_ & z  & \_ & \, \\
(\alpha_2 \alpha_3),\:\: \mu_1 &\mbox{d1:} \,\, & \_ & \_ & \_ & \_ & \_ & \psi & \, \\
(\alpha_1 \alpha_3),\:\: \mu_2 &\mbox{d5:} \,\, & 1  & 2  & 3  & 4  & \_ & \psi & \, \\
(\alpha_1 \alpha_2),\:\: \mu_3 &\mbox{kkm:}\,\, & 1  & 2  & 3  & 4  & (z) & \psi & \, .
\end{array}
\eeqa
We will show that by appropriately choosing the parameters of the
solution we can manage to eliminate the global pathologies produced by
having boosted the Kaluza-Klein monopoles along their fiber directions.
Roughly speaking, the pathology from boosting the kkm dipole induced by
charging up the solution is cancelled against the pathologies from
boosting the dipoles of the initial configuration \reef{seedarray}. To
this effect, a single dipole in the seed solution would be sufficient,
but using the complete solution \reef{seedarray} we will obtain a larger
family of charged black rings.

There is a more symmetrical M-theory version of the solutions, obtained
by performing $\to T(34)\to T(z)$ and then uplifting to eleven
dimensions. This configuration is
\beqa
\begin{array}{clcccccccl}
\alpha_1 &\mbox{M2:} \,\, & 1  & 2  & \_  & \_  & \_ & \_ & \_ & \, \\
\alpha_2 &\mbox{M2:} \,\, & \_ & \_ & 3 & 4 & \_ & \_ & \_ & \, \\
\alpha_3 &\mbox{M2:} \,\, & \_ & \_ & \_ & \_ & 5 & 6 & \_ & \, \\
(\alpha_2 \alpha_3),\:\: \mu_1 &\mbox{m5:} \,\, & \_ & \_ & 3 & 4 & 5 & 6 & \psi & \, \\
(\alpha_1 \alpha_3),\:\: \mu_2 &\mbox{m5:} \,\, & 1  & 2  &\_ &\_ & 5 & 6 & \psi & \, \\
(\alpha_1 \alpha_2),\:\: \mu_3 &\mbox{m5:} \,\, & 1  & 2  & 3 & 4 &\_
&\_ & \psi & \, ,
\end{array}
\label{Marray}\eeqa
where we have set $z_5 \equiv z$, and the eleventh dimensional direction is
$z_6$. It is this form of the solution that we present next.

\subsection{Solution}
\label{subsec:3chsol}
The metric for the eleven-dimensional solution is
\beqa
ds_\rom{11D}^2&=& ds_\rom{5D}^2
+\left[\frac{1}{h_1}\, \frac{H_1(y)}{H_1(x)}\right]^{2/3}
\left[h_2\, h_3\, \frac{H_2(x)H_3(x)}{H_2(y)H_3(y)}\right]^{1/3}
(dz_1^2+dz_2^2)\nonumber\\[2mm] \label{eqn:11d}
&& \hspace{1.cm}
+\left[\frac{1}{h_2}\, \frac{H_2(y)}{H_2(x)}\right]^{2/3}
\left[h_1\, h_3\, \frac{H_1(x)H_3(x)}{H_1(y)H_3(y)}\right]^{1/3}
(dz_3^2+dz_4^2) \\[2mm]
&& \hspace{1.cm}\nonumber
+\left[\frac{1}{h_3}\, \frac{H_3(y)}{H_3(x)}\right]^{2/3}
\left[h_1\, h_2\, \frac{H_1(x)H_2(x)}{H_1(y)H_2(y)}\right]^{1/3}
(dz_5^2+dz_6^2)  \, ,
\eeqa
where
\beqa
  \label{5dmetric}
  ds_\rom{5D}^2 &=& -\frac{1}{(h_1 h_2 h_3)^{2/3}}
\frac{H(x)}{H(y)}
\frac{F(y)}{F(x)}
\Big(dt+\omega_\psi(y)d\psi
+\omega_\phi(x)d\phi\Big)^2\nonumber\\[3mm]
&\;&
+(h_1 h_2 h_3)^{1/3} F(x) H(x) H(y)^2
\\[3mm]\nonumber
&&\hspace{1cm}
\times
\frac{R^2}{(x-y)^2}\Bigg[
     - \frac{G(y)}{F(y) H(y)^3} \, d\psi^2
     - \frac{dy^2}{G(y)}
     + \frac{dx^2}{G(x)}
     + \frac{G(x)}{F(x) H(x)^3} \, d\phi^2
   \Bigg] \, ,
\eeqa
and the three-form potential is
\beq
  \mathcal{A} =
    A^1 \wedge dz_1 \wedge dz_2
  + A^2 \wedge dz_3 \wedge dz_4
  + A^3 \wedge dz_5 \wedge dz_6 \, .
\label{threepot}\eeq
The explicit expressions for the components of the one-forms $A^i$
($i=1,2,3$) are given in appendix \ref{appA}.

We have defined the
following functions
\beq
  \begin{array}{lclclcl}
  F(\xi) &=& 1 + \lambda \xi \, ,&~~~~~~~&
  G(\xi) &=& (1-\xi^2)(1+\nu \xi)\, , ~~~~~~\\[2mm]
  H_i(\xi) &=& 1- \mu_i \xi \, ,&~~~~~~&
  H(\xi) &\equiv& \big[ H_1(\xi)H_2(\xi)H_3(\xi) \big]^{1/3}\, ,
  \end{array}
\label{FGH}\eeq
and
\beq
  h_i =  {{c_i}}^2 - U_i  {{s_i}}^2\, ,
\eeq
where the functions $U_i$ are defined in \reef{Ui} and, in order to reduce
notational clutter, we have introduced
\beq
c_i\equiv \cosh\alpha_i\,,\qquad s_i\equiv \sinh\alpha_i\,.
\eeq
It is useful to give explicit expressions for the $h_i$:
\beqa
  h_1 &=& 1 + \frac{H_1(y) {s^2_1}}{H_1(x)F(x)H(y)^3}
  \bigg[
  (\lambda - \mu_1 + \mu_2 + \mu_3)(x-y)
        -(\mu_2 \mu_3 +\lambda \mu_1)(x^2-y^2)\nonumber \\
  && \hspace{4.5cm}
        +(\mu_1 \mu_2 \mu_3 + \lambda \mu_1 \mu_2
          + \lambda \mu_1 \mu_3 - \lambda \mu_2 \mu_3) \,xy (x-y)
  \bigg]\, ,
\eeqa
and $h_2$, $h_3$ obtained by exchanging $1\leftrightarrow 2$ and
$1\leftrightarrow 3$, respectively.

The components of the one-form $\omega=\omega_\psi\; d\psi+\omega_\phi\;
d\phi$ are
\beqa
  \omega_\psi(y) &=& R(1+y)
  \bigg[
    \frac{C_\lambda}{F(y)}  {c_1}  {c_2}  {c_3}
   - \frac{C_1}{H_1(y)}  {c_1}  {s_2}  {s_3}
   - \frac{C_2}{H_2(y)}  {s_1}  {c_2}  {s_3}
   -\frac{C_3}{H_3(y)}  {s_1}  {s_2}  {c_3}
  \bigg]\,,
     \\[2mm]
  \omega_\phi(x) &=& - R(1+x)
  \bigg[
    \frac{C_\lambda}{F(x)}  {s_1}  {s_2}  {s_3}
- \frac{C_1}{H_1(x)}  {s_1}  {c_2}  {c_3}
   - \frac{C_2}{H_2(x)}  {c_1}  {s_2}  {c_3}
   -\frac{C_3}{H_3(x)}  {c_1}  {c_2}  {s_3}
  \bigg] \, ,
\eeqa
where
\beq
\label{Cs}
  C_\lambda = \epsilon_\lambda \sqrt{\lambda (\lambda - \nu)
  \frac{1+\lambda}{1-\lambda}} \, ,~~~~~~
  C_i = \epsilon_i \sqrt{\mu_i (\mu_i + \nu)
  \frac{1-\mu_i}{1+\mu_i}}
\eeq
for $i=1,2,3$.
A choice of sign $\epsilon_i, \epsilon_\lambda = \pm 1$ has been
included explicitly. 

We assume that the coordinates $z_i$,
$i=1,\dots,6$ are periodically identified.
The coordinates $x$ and $y$ take values in the ranges
\beq
  -1 \le x \le 1 \, ,~~~~~~
  -\infty < y \le -1  \, ,~~~~~~
  \frac{1}{\min{\mu_i}} < y < \infty  \, .
\eeq
The solution has three Killing vectors, $\partial_t$,
$\partial_\psi$, and $\partial_\phi$, and is characterized by eight
dimensionless parameters
$\lambda, \nu,
\mu_i,\alpha_i$, plus the scale parameter $R$, which has dimension of length.

Without loss of generality
we can take $R>0$.
The parameters $\lambda, \nu, \mu_i$ are
restricted as
\beq
  0 < \nu \le \lambda < 1 \, ,~~~~~~
  0 \le \mu_i < 1 \, ,
\eeq
while the $\alpha_i$ can initially take any real value. These ranges of values are
typically sufficient to avoid the appearance of naked curvature
singularities. Below we will discuss how the elimination of other pathologies
will reduce the total number of free parameters from nine to
seven.

Each $\alpha_i$ is associated with an
M2-brane charge; taking $\alpha_i=0$ sets the corresponding M2-brane
charge to zero.
In particular, taking all $\alpha_i=0$, we recover the
dipole black rings of \cite{RE}.
The solutions contain contributions to the M5-brane dipole charges
that originate both from the parameters $\mu_i$ as well as from the
boosts, see \reef{Marray}. The precise relation between the parameters
and the charges will be given below.

Asymptotic infinity is at $x, y \to -1$.
Since $x=-1$ and $y=-1$ are fixed
point sets of respectively $\partial_\phi$ and $\partial_\psi$,
the periodicites of $\psi$ and $\phi$ must be chosen so as to avoid conical
defects that would extend to
infinity. The required periodicities are
\beq
  \Delta\psi = \Delta\phi
  = 2 \pi \frac{\sqrt{1-\lambda}}{1-\nu}
    \prod_{i=1}^3 \sqrt{1+\mu_i} \, .
\label{delphipsi}\eeq
Defining canonical angular variables
\beq
\tilde{\psi} = \frac{2\pi}{\Delta\psi}\, \psi\,,\qquad
\tilde{\phi} = \frac{2\pi}{\Delta\phi}\, \phi\,,
\label{canang}\eeq 
and performing the coordinate transformation
\beq
  \zeta_1 = \tilde{R} \frac{\sqrt{-1-y}}{x-y}\, ,~~~~~~~
  \zeta_2 = \tilde{R} \frac{\sqrt{1+x}}{x-y}\, ,~~~~~~~
  \tilde{R}^2 = 2 R^2\, \frac{1-\lambda}{1-\nu}
                    \prod_{i=1}^3 (1+\mu_i) \, ,
\eeq
the five-dimensional asymptotic metric takes the manifestly flat form
\beq
  ds_\rom{5D}^2 = - dt^2 + d\zeta_1^2 + \zeta_1^2 \, d\tilde{\psi}^2
     + d\zeta_2^2 + \zeta_2^2 \, d\tilde{\phi}^2 \, .
\eeq
Thus the metric is asymptotically five-dimensional Minkowski space
times a six-torus.

\subsubsection*{Removing Dirac-Misner strings}
We are interested in ring-like solutions with horizon topology
$S^1\times S^2$.
In order that $(x,\phi)$ parameterize a two-sphere,
$\partial_\phi$ must have fixed-points at $x=\pm 1$, corresponding to
the poles of the $S^2$.
However, note that when the three $\alpha_i$ are non-zero, the orbit of
$\partial_\phi$
does not close off at $x=1$, since $\omega_\phi(x=1) \ne 0$. This
can be interpreted as the presence of Dirac-Misner strings --- a
geometric analogue of
Dirac strings discussed by Misner in the Taub-NUT solution \cite{misner}.
Their
appearance in this
solution can be traced to the fact that, to obtain it, we have boosted
along the fiber of a Kaluza-Klein
monopole \cite{EE}. By analogy with the Dirac monopoles, one might try
to eliminate the strings by covering the geometry with
two patches, each one regular at each pole. However, the
coordinate transformation in the region where the two patches overlap
would require $t$ to be periodically identified with
period $\Delta t= \omega_\phi(x=1)\Delta\phi$ (or an integer fraction of
this). Closed timelike curves would then be present everywhere outside
the horizon.

To remove the pathology we must therefore require that the form
$\omega$ be globally well-defined, \ie that $\omega_\phi(x=\pm 1) =
0$. This places a constraint on the parameters of the solution of the
form
\beqa
  &&\frac{C_\lambda}{1+\lambda}
   {s_1}  {s_2}  {s_3}=
  \frac{C_1}{1-\mu_1}
   {s_1}  {c_2}  {c_3}
  +\frac{C_2}{1-\mu_2}
   {c_1}  {s_2}  {c_3}
  +\frac{C_3}{1-\mu_3}
   {c_1}  {c_2}  {s_3}
   \, .
  \label{DMcond}
\eeqa
Imposing this condition, $\omega_\phi$ can be written as
\beq
  \omega_\phi(x) = - \frac{R (1-x^2)}{F(x)}
  \Bigg[~
    \frac{\lambda+\mu_1}{1-\mu_1}
    \frac{C_1}{H_1(x)}
     {s_1}  {c_2}  {c_3}
    +\frac{\lambda+\mu_2}{1-\mu_2}
    \frac{C_2}{H_2(x)}
     {c_1}  {s_2}  {c_3} \\[2mm]
    +\frac{\lambda+\mu_3}{1-\mu_3}
    \frac{C_3}{H_3(x)}
     {c_1}  {c_2}  {s_3}
  \Bigg] \, ,
\eeq
which is manifestly regular.

\subsubsection*{Balancing the ring}

The choice \reef{delphipsi} for the period of $\phi$ makes the orbits of
$\partial_\phi$ close off smoothly at $x=-1$. We have also required that
the orbits of $\partial_\phi$ close off at the other pole, $x=+1$, by
imposing the condition \reef{DMcond}, but there still remains the
possibility that conical defects are present at this pole. Smoothness at
$x=1$ requires another, specific value for $\Delta\phi$, and it is easy
to see that this is compatible with \reef{delphipsi} only if the
parameters satisfy the equation
\beq
  \label{balance}
  \left(\frac{1-\nu}{1+\nu}\right)^2
  =
  \frac{1-\lambda}{1+\lambda}
  \prod_{i=1}^3 \frac{1+\mu_i}{1-\mu_i} \, .
\eeq
Violating this condition results in a disk-like conical singularity
inside the ring at $x=1$. Depending on whether there is an excess or
deficit angle, the disk provides a push or pull to keep the ring in
equilibrium. Thus \reef{balance} is a balancing condition. We assume the
ring is balanced,
\ie that \reef{balance} holds. This condition is independent of
$\alpha_i$ and hence the same condition was found for the dipole black
rings in \cite{RE}.

\bigskip

With the balancing condition \reef{balance} and the Dirac-Misner
condition \reef{DMcond}, the solution contains seven independent
parameters: the scale $R$, plus six dimensionless parameters. These may
be taken to be $\mu_i$ and $\alpha_i$, if $\lambda$ and $\nu$ are
eliminated through \reef{DMcond} and \reef{balance}.

\subsection{Properties}

We give here expressions for the conserved charges (mass, angular
momentum and net charge) as well as for the dipole charges. We then
analyze the horizon geometry and compute the horizon area, temperature
and angular velocity of the black rings.

\subsubsection*{Asymptotic charges}

If we assume that the $z_i$ directions are all compact with period
$2\pi \ell$ then the five-dimensional Newton's constant is related to
the 11D coupling constant $\kappa$ through $\kappa^2=8\pi G_5(2\pi
\ell)^6$. Also, note that the six-torus parametrized by the $z_i$ has
constant volume. This constraint implies that the five-dimensional
metric $ds_\rom{5D}^2$ is the same as the Einstein-frame metric arising
from the reduction of the eleven-dimensional metric \reef{eqn:11d} on
the $T^6$. The mass and angular momenta in five dimensions can then be
obtained from the asymptotic form of the metric.

The mass is most simply expressed as
\beq
M=\frac{\pi}{4G_5}\Big(Q_1 \coth 2\alpha_1+Q_2 \coth 2\alpha_2+Q_3
\coth 2\alpha_3\Big)\,,
\label{mass}\eeq
in terms of the M2-brane charges carried by the solution,
\beqa
\label{Qs}
Q_1&=&\frac{R^2 \sinh{2\alpha_1}}{1-\nu}
\Big[
  \lambda-\mu_1+\mu_2+\mu_3
  +2(\mu_2\mu_3+\lambda\mu_1)
  +\lambda(\mu_1\mu_2+\mu_1\mu_3-\mu_2\mu_3)+\mu_1\mu_2\mu_3
\Big]\;, \nonumber\\
Q_2&=&\frac{R^2 \sinh{2\alpha_2}}{1-\nu}
\Big[
  \lambda+\mu_1-\mu_2+\mu_3
  +2(\mu_1\mu_3+\lambda\mu_2)
  +\lambda(\mu_1\mu_2-\mu_1\mu_3+\mu_2\mu_3)+\mu_1\mu_2\mu_3
\Big]\;, \nonumber\\
Q_3&=&\frac{R^2 \sinh{2\alpha_3}}{1-\nu}
\Big[
  \lambda+\mu_1+\mu_2-\mu_3
  +2(\mu_1\mu_2+\lambda\mu_3)
  +\lambda(-\mu_1\mu_2+\mu_1\mu_3+\mu_2\mu_3)+\mu_1\mu_2\mu_3
\Big]   \;, \nonumber\\
\eeqa
and satisfies the BPS bound
\beq
M\geq  \frac{\pi}{4G_5}\Big(\,|Q_1|+|Q_2|+|Q_3|\,\Big)\,.
\label{bpsbound}
\eeq
The two angular momenta are
\beqa
J_{\psi}\!\!&=&\!\!\frac{\pi R^3}{2G_5}
\frac{(1-\lambda)^{3/2}}{(1-\nu)^2}
\Bigg[
\prod_{i=1}^3(1+\mu_i)^{3/2}
\Bigg]
\Bigg[\;
  \frac{C_{\lambda}}{1-\lambda}
   {c_1}  {c_2}  {c_3}
-\frac{C_1}{1+\mu_1} {c_1}  {s_2}  {s_3}
-\frac{C_2}{1+\mu_2} {s_1}  {c_2}  {s_3}
-\frac{C_3}{1+\mu_3} {s_1}  {s_2}  {c_3}
\;\Bigg] ,
\nonumber\\
\\[3mm]
J_{\phi}\!\!&=&\!\!-\frac{\pi R^3}{2G_5}
\frac{(1-\lambda)^{3/2}}{(1-\nu)^2}
\Bigg[
\prod_{i=1}^3(1+\mu_i)^{3/2}
\Bigg]
\Bigg[\;
  \frac{C_{\lambda}}{1-\lambda}
   {s_1}  {s_2}  {s_3}
  -\frac{C_1}{1+\mu_1}
   {s_1}  {c_2}  {c_3}
  -\frac{C_2}{1+\mu_2}
   {c_1}  {s_2}  {c_3}
  -\frac{C_3}{1+\mu_3}
   {c_1}  {c_2}  {s_3}
\;\Bigg] .
\nonumber\\
\eeqa
Eq.~\reef{DMcond} can be used to write the latter as
\beqa
J_{\phi}\!\!&=&\!\!-\frac{\pi R^3}{G_5}
\frac{\sqrt{1-\lambda}}{(1-\nu)^2}
\Bigg[
\prod_{i=1}^3(1+\mu_i)^{3/2} 
\Bigg]
\Bigg[\;
  \frac{\lambda+\mu_1}{1-\mu_1^2}\;C_1
   {s_1}  {c_2}  {c_3}
+\frac{\lambda+\mu_2}{1-\mu_2^2}\;C_2
   {c_1}  {s_2}  {c_3}
  +\frac{\lambda+\mu_3}{1-\mu_3^2}\;C_3
   {c_1}  {c_2}  {s_3}
\;\Bigg] \;,\nonumber\\
\eeqa
although it does not lead to any simpler expressions for
$J_\psi$ or $Q_i$.

\subsubsection*{Dipole charges}

The dipole charges are given by
\beq
\label{qs}
  q_i=\frac{1}{2\pi (2\pi \ell)^2} \int_{S^2 \times T^2} d\mathcal{A}
  = \frac{1}{2\pi} \int_{S^2} dA^i
  = \frac{\Delta\phi}{2\pi}\left[A^i_\phi(x=1)-A^i_\phi(x=-1) \right]\, ,
\eeq
where the two-sphere parameterized by $(x,\phi)$ surrounds a
constant-$\psi$ slice of the black
ring and for $i=1,2,3$ the two-torus is parameterized by $z_1$-$z_2$,
$z_3$-$z_4$ or $z_5$-$z_6$, respectively.
For generic values of the parameters the dipole charges are not
well-defined since the
expressions \reef{qs} are $y$-dependent. The condition for the corresponding gauge
fields to be well-defined is the same as imposing the absence of
Dirac-Misner strings \reef{DMcond}. With this, the $y$-dependence drops
out and we
find
\beqa
 q_1&=&-\frac{2R}{{s_1}}\frac{\sqrt{1-\lambda}}{1-\nu}
    \Bigg[\prod_{i=1}^3 \sqrt{1+\mu_i}\Bigg]\left[
 \frac{C_2}{1-\mu_2} {s_2} {c_3}
 +\frac{C_3}{1-\mu_3} {c_2}  {s_3}\right]\;,\\
 q_2&=&-\frac{2R}{{s_2}}\frac{\sqrt{1-\lambda}}{1-\nu}
    \Bigg[\prod_{i=1}^3 \sqrt{1+\mu_i}\Bigg]\left[
 \frac{C_1}{1-\mu_1} {s_1} {c_3}
 +\frac{C_3}{1-\mu_3} {c_1}  {s_3}\right]\;,\\
 q_3&=&-\frac{2R}{{s_3}}\frac{\sqrt{1-\lambda}}{1-\nu}
    \Bigg[\prod_{i=1}^3 \sqrt{1+\mu_i}\Bigg]\left[
 \frac{C_2}{1-\mu_2} {s_2} {c_1}
 +\frac{C_1}{1-\mu_1} {c_2}  {s_1}\right]\;.\label{q3}
\eeqa
One can easily verify that
\beq
J_\phi=\frac{\pi}{8G_5}\left(q_1Q_1+q_2Q_2+q_3Q_3\right)\,.
\label{JqQ}\eeq
This identity reflects the fact that the second angular momentum
$J_\phi$ appears as a result of charging up the dipole rings.

\subsubsection*{Non-uniqueness}

There are seven parameters in the solution, but only six conserved
charges at infinity, ($M$, $J_\psi$, $J_\phi$, $Q_{1,2,3}$). So fixing
these parameters we can expect to find a one-parameter continuous
non-uniqueness.

\subsubsection*{Horizon}

As for the dipole black rings of \cite{RE}, we expect the event horizon to be
located at $y=y_h\equiv -1/\nu$. At $y=y_h$, $g_{yy}$ blows up, but this is
just a coordinate singularity which can be removed by the coordinate
transformation $(t,\psi) \to (v,\psi')$ given as
\beq
  dt = dv
   + \omega_\psi(y)\,\frac{\sqrt{-F(y)H(y)^3}}{G(y)} dy \, ,
  \hspace{1cm}
  d\psi = d\psi' - \frac{\sqrt{-F(y)H(y)^3}}{G(y)} dy \, .
  \label{horcoord}
\eeq
Then the five-dimensional part of the metric is
\beqa
ds_\rom{5D}^2&=&-\frac{1}{(h_1 h_2 h_3)^{2/3}}
\frac{H(x)}{H(y)}
\frac{F(y)}{F(x)}
\Big(dv+\omega_\psi(y)d\psi'+\omega_\phi(x)d\phi\Big)^2\nonumber\\[3mm]
&\;&
+(h_1 h_2 h_3)^{1/3}
 H(x) H(y)^2 F(x)
\nonumber\\[3mm]
&&\hspace{1cm}
\times
\frac{R^2}{(x-y)^2}\bigg[
     - \frac{G(y)}{F(y) H(y)^3} \, d\psi'^2
     - \frac{2\, d\psi' dy }{\sqrt{-F(y)H(y)^3}}
     + \frac{dx^2}{G(x)}
     + \frac{G(x)}{F(x) H(x)^3} \, d\phi^2
   \bigg] \, ,\nonumber\\
\eeqa
and thus the full metric is manifestly regular at $y=y_h$.

The metric on (a spatial section of) the horizon is
\beqa
\label{hormet}
ds_{H}^2 &=&
\frac{1}{(h_1 h_2 h_3)^{2/3}}
\frac{H(x)}{F(x)}
\frac{|F(y_h)|}{H(y_h)}
\Big(\omega_\psi(y_h)d\psi'+\omega_\phi(x)d\phi\Big)^2\nonumber\\[3mm]
&\;&
+(h_1 h_2 h_3)^{1/3}
 H(x) H(y_h)^2 F(x)
\frac{R^2}{(x-y_h)^2}\left[ \frac{dx^2}{G(x)}
     + \frac{G(x)}{F(x) H(x)^3} \, d\phi^2
   \right] \, ,
\eeqa
where the $h_i$ are evaluated at $y_h$, but recall
that they also depend on $x$.

In order to better understand the geometry of this horizon, let us
consider first the following simpler metric,
\beqa
\label{toyhor}
ds^2&=&R_1^2\left(d\psi'+ k(1-x^2)d\phi\right)^2+R_2^2\left(\frac{dx^2}{1-
x^2}+(1-x^2)d\phi^2\right)\nonumber\\
&=&R_1^2\left(d\psi'+ k\sin^2\theta
d\phi\right)^2+R_2^2\left(d\theta^2+\sin^2\theta d\phi^2\right)\,,
\eeqa
where the second expression is obtained by making $x=\cos\theta$, and
$R_1$, $R_2$, $k$, are constants. This
is topologically $S^1\times S^2$. Due to the cross-term
$g_{\psi'\phi}$ the product is twisted\footnote{Note that at constant $\theta\neq
0,\pi$ one recognizes a twisted 2-torus.}, but, since
$g_{\psi'\phi}$ vanishes at
$\theta=0,\pi$, \ie at $x=\pm 1$, the fibration of the $S^1$ over the
$S^2$ is topologically trivial and globally well-defined. 

The horizon metric \reef{hormet} describes a geometry topologically
equivalent to \reef{toyhor} (recall that $G(x), \omega_\phi(x) \propto
(1-x^2)$), but now $R_1$, $R_2$, $k$ are functions of the polar
coordinate $x\in[-1,1]$, everywhere regular and non-vanishing. So the
horizon of these black rings is topologically $S^1\times S^2$, but the
radii of the $S^1$ and $S^2$, and the twisting, are not constant but
change with the latitude of the $S^2$.

The horizon area is
\beqa
  \mathcal{A}_H &=& 8 \pi^2 R^3 \,
  \frac{(1-\lambda)(\lambda - \nu)^{1/2}}{(1-\nu)^2(1+\nu)}
  \Bigg[\prod_{i=1}^3 (1+\mu_i) (\nu+ \mu_i)^{1/2}\Bigg]
   \nonumber \\
  && \hspace{1cm}
  \times\Bigg|
  \frac{C_\lambda}{\lambda-\nu} {c_1} {c_2} {c_3}
  +\frac{C_1}{\nu+\mu_1} {c_1} {s_2} {s_3}
  +\frac{C_2}{\nu+\mu_2} {s_1} {c_2} {s_3}
  +\frac{C_3}{\nu+\mu_3} {s_1} {s_2} {c_3}
  \Bigg| \, .
\eeqa

When all $\mu_i$ are nonzero, there is also an inner
horizon at $y=-\infty$.
At $y=1/\min{\mu_i}$ there is a curvature singularity hidden behind
the horizons. The Killing vector $\partial_t$ becomes
spacelike at $y=-1/\lambda$, so $-1/\nu < y < -1/\lambda$ is the
ergoregion. The ergosurface at $y=-1/\lambda$ has topology $S^1 \times
S^2$.

The horizon is generated by the orbits of the Killing vector
{\boldmath$\xi$}$=\partial/\partial
t-\Omega_{\psi}\,\partial/\partial{\tilde{\psi}}$ (where $\tilde\psi$ is
the angle in \reef{canang}), with the angular velocity of the
horizon given by
\beqa
\Omega_\psi^{-1}&=&\frac{\Delta\psi}{2\pi}\;\omega_\psi(y_h)\\
&=&R\sqrt{1-\lambda}\Bigg[\prod_i\sqrt{1+\mu_i}\Bigg] \Bigg|
  \frac{C_\lambda}{\lambda-\nu} {c_1} {c_2} {c_3}
  +\frac{C_1}{\nu+\mu_1} {c_1} {s_2} {s_3}
  +\frac{C_2}{\nu+\mu_2} {s_1} {c_2} {s_3}
  +\frac{C_3}{\nu+\mu_3} {s_1} {s_2} {c_3}
  \Bigg|\,.\nonumber
\eeqa
Note that the angular velocity in the $\phi$ direction vanishes even if
$J_\phi\neq 0$, which is rather unusual for a non-supersymmetric
solution.

The temperature, obtained from the surface gravity at the horizon, is
\beqa
T_H^{-1}=4\pi R\frac{\sqrt{\lambda-\nu}\prod_i\sqrt{\mu_i+\nu}}{\nu(1+\nu)}
\Bigg|
  \frac{C_\lambda}{\lambda-\nu} {c_1} {c_2} {c_3}
  +\frac{C_1}{\nu+\mu_1} {c_1} {s_2} {s_3}
  +\frac{C_2}{\nu+\mu_2} {s_1} {c_2} {s_3}
  +\frac{C_3}{\nu+\mu_3} {s_1} {s_2} {c_3}
  \Bigg|.\nonumber\\
\eeqa
The entropy of the black ring is $S = \mathcal{A}_H/4$. We note that
$T_H S$ is quite simple, but $\Omega_\psi J_\psi$ is not. It will be
interesting to understand the thermodynamics of these non-supersymmetric
black rings. The first law for black rings with dipole charges is currently being
investigated \cite{CH}.

\setcounter{equation}{0}
\section{Extremal and supersymmetric limits}
\label{sec:extsusy}

In order to avoid possible confusion, it may be worth recalling that the
extremal limit and the supersymmetric limit of a black hole solution,
even if they often coincide, in general need not be the same. The
extremal limit is defined as the limit where the inner and outer
horizons of a black hole coincide. Then, if the horizon remains regular,
its surface gravity, and hence its temperature, vanish. The
supersymmetric limit, instead, is one where the limiting solution
preserves a fraction of supersymmetry and saturates a BPS bound. If the
horizon remains regular in the supersymmetric limit, it must be
degenerate, hence extremal. But the converse is not true in general. The
extremal limit of a black hole need not be supersymmetric. A familiar
example of this is the Kerr-Newman solution, whose extremal limit at
maximal rotation is not supersymmetric, and whose supersymmetric limit,
with $M=|Q|$, cannot have a regular horizon at finite rotation.

Let us first discuss the extremal limit, since it is simpler. If
$\nu=0$, and all the $\mu_i$ are non-zero, the inner and outer horizon
coincide and the ring has a degenerate horizon with zero temperature. So
the extremal limit is $\nu\to 0$. One finds a regular horizon of {\it
finite area} as long as all $\mu_i$ are non-vanishing. Such extremal
solutions have proven useful in order to understand the microphysics of
black rings \cite{RE}. However, for finite values of the parameters
other than $\nu$, these extremal solutions are not supersymmetric. This
is clear from the fact that they do not saturate the BPS bound
\reef{bpsbound}.

In order to saturate this bound, we see from \reef{mass} that we must
take $|\alpha_i|\to\infty$.
But before analyzing this limit, it is important to realize that the
solutions in this paper cannot reproduce the most general
supersymmetric rings with three charges and three dipoles in
ref.~\cite{EEMR2}. A simple way to see this is by noting that the latter have
\beq
J_\phi=\frac{\pi}{8G_5}\left(q_1Q_1+q_2Q_2+q_3Q_3-q_1q_2q_3\right)
\quad\mathrm{(BPS\,\,ring)}\,,
\eeq
instead of \reef{JqQ}. The parameter count mentioned in the
introduction also leads to this conclusion. Furthermore, for the
supersymmetric solutions with all three charges and three dipoles, each
of the functions $\omega_\psi$ and $\omega_\phi$ depend on {\em both}
$x$ and $y$, whereas here we have $\omega_\psi(y)$ and
$\omega_\phi(x)$.

It follows that at most we can recover a supersymmetric ring with three
charges and {\em two} dipoles. This ring does not have a regular
horizon. We believe, however, that this limitation of the
non-supersymmetric solutions is not fundamental, but instead is just a
shortcoming of our construction starting from the seed in \cite{RE} (so
far the most general seed solution available). We expect that a more
general non-supersymmetric black ring solution which retains the three
charges and the three dipoles in the supersymmetric limit exists.

In order to take the supersymmetric limit in such a way that three
charges and two dipoles survive,
take $\alpha_1, \alpha_2, \alpha_3 \to
\infty$ and $\lambda, \nu, \mu_i \to 0$ such that
$e^{2\alpha_1} \sim e^{2\alpha_2} \sim e^{\alpha_3}$ and
$\lambda \sim \mu_3 \sim e^{-\alpha_3}$, while
$\mu_1 \sim \mu_2 \sim \nu \sim (\lambda - \mu_3) \sim
e^{-2\alpha_3}$. Note the latter implies that $\lambda = \mu_3
+ O(e^{-2\alpha_3})$, which we shall use in the following.

These scalings are conveniently encoded by saying that in the limit we
keep fixed the following quantities:
\beqa
  &&
  \lambda e^{2\alpha_1} = \frac{Q_1}{R^2} \, ,~~~~~~~
  \lambda e^{2\alpha_2} = \frac{Q_2}{R^2}  \, ,~~~~~~~
  \frac{1}{2}(\lambda + \mu_1 + \mu_2 - \mu_3 + 2 \lambda \mu_3)e^{2\alpha_3}
  = \frac{Q_3}{R^2}\, ,\\
  &&
  -\epsilon_3 \lambda e^{\alpha_3+\alpha_2-\alpha_1}=\frac{q_1}{R}\, ,~~~~~~~
  -\epsilon_3 \lambda e^{\alpha_3-\alpha_2+\alpha_1}=\frac{q_2}{R}\, ,~~~~~~~\\
  &&
  (\mu_1 + \mu_2 + \nu) e^{2\alpha_3} = \frac{a^2}{R^2} \label{a2}\, ,~~~~~~~\\
  &&
  \frac{1}{2}\left[\mu_1 + \mu_2 + \nu+\epsilon_\lambda
  \left(
    \epsilon_1 \sqrt{\mu_1(\nu+\mu_1)}
    +\epsilon_2 \sqrt{\mu_2(\nu+\mu_2)}
  \right)\right] e^{2\alpha_3} = \frac{b^2}{R^2} \label{b2}\, .~~~~~~~
\eeqa
The $Q_i$ and $q_i$ are actually the limits of the charges and dipoles
in \reef{Qs} and \reef{qs}.
Note that now the limiting $q_{1,2}$ and $Q_{1,2}$ are not independent, but
satisfy
\beq
  q_1 Q_1 = q_2 Q_2 \, .
  \label{q1q2}
\eeq

Recall that the $\epsilon_\lambda,\epsilon_i$ are choices of signs, $\epsilon_\lambda,\epsilon_i=\pm 1$.
We have arbitrarily chosen the boosts $\alpha_i$ to be positive. This
then requires the sign choice $\epsilon_\lambda = \epsilon_3$ in order
that the cancellation of Dirac-Misner strings \reef{DMcond} be possible.
The parameters $a^2$ and $b^2$ are non-negative numbers, $a^2\geq b^2\geq
0$ (for any choice of the signs $\epsilon_\lambda$, $\epsilon_i$), and
we shall presently see that after imposing the balancing condition and
the Dirac-Misner condition, they drop out from the solution. 

We demand that the supersymmetric limit is approached
through a sequence of black rings which are regular (on and outside the
horizon), and therefore require that they satisfy the balancing
condition \reef{balance}. In the limit, this becomes
\beq
  2\nu = \lambda - \mu_1 - \mu_2 - \mu_3  \, ,
  \label{balance3}
\eeq
which can be written
\beq
  \label{balanceextr}
  Q_3 - q_1 q_2 = a^2 \, .
\eeq
The Dirac-Misner condition \reef{DMcond} gives
\beq
  R^2 q_1 q_2 (Q_1 + Q_2) = Q_1 Q_2 (Q_3 - q_1 q_2 - b^2) \, .
  \label{DMextr}
\eeq

Now turning to the solution, we find in the supersymmetric limit that
\beq
  h_1 = 1 + \frac{Q_1}{2R^2} (x-y) \, ,~~~~~
  h_2 = 1 + \frac{Q_2}{2R^2} (x-y) \, ,~~~~~
  h_3 = 1 + \frac{Q_3-q_1 q_2}{2R^2} (x-y)
  - \frac{q_1 q_2}{4R^2}(x^2-y^2)  \, .
\eeq
Also, using the balancing condition \reef{balance3},
the Dirac-Misner condition \reef{DMextr}, and
\reef{q1q2} to rewrite the expressions, we get
\beqar
  \omega_\psi(y) &=& \epsilon_\lambda
  \left[
    \frac{1}{2}(q_1 + q_2)(1+y)
    - \frac{1}{8R^2} (y^2-1) (q_1 Q_1 + q_2 Q_2)
  \right] \, ,\\
  \omega_\phi(x) &=& -\epsilon_\lambda
    \frac{1}{8R^2} (1-x^2) (q_1 Q_1 + q_2 Q_2)
  \,.
\eeqar
The supersymmetric-limit metric is
\beqa
ds_{11}^2&=&-\frac{1}{(h_1 h_2 h_3)^{2/3}}
\left[dt+\omega_\psi(y)d\psi+\omega_\phi(x)d\phi\right]^2\nonumber\\[3mm]
&\;&
+(h_1 h_2 h_3)^{1/3}
\Bigg\{
\frac{R^2}{(x-y)^2}\bigg[
     (y^2-1)\, d\psi^2
     + \frac{dy^2}{y^2-1}
     + \frac{dx^2}{1-x^2}
     + (1-x^2) \, d\phi^2
   \bigg]
   \nonumber\\[3mm]
&\;&\hspace{1.6cm}
+\frac{1}{h_1}(dz_1^2+dz_2^2)
+\frac{1}{h_2}(dz_3^2+dz_4^2)
+\frac{1}{h_3}(dz_5^2+dz_6^2)
\Bigg\} \, .
\eeqa
For the three-form potentials (given in appendix \ref{appA}), we find after imposing the
balancing condition \reef{balance3} and the Dirac-Misner condition
\reef{DMextr2}
\beq
A^i_t=h_i^{-1}-1,\qquad i=1,2,3,
\eeq
while
\beq
  A^i_\psi = \frac{1}{h_i} \omega_\psi - \frac{q_i}{2}(1+y)\, ,
  ~~~~~
  A^i_\phi = \frac{1}{h_i} \omega_\phi - \frac{q_i}{2}(1+x)\, ,
\qquad i=1,2,
\eeq
and
\beq
  A^3_\psi  = \frac{1}{h_3}  \omega_\psi\, ,~~~~~~~~
  A^3_\phi  = \frac{1}{h_3}  \omega_\phi \,.
\eeq
Choosing $\epsilon_\lambda = +1$, this matches exactly the
full eleven-dimensional supersymmetric solution of \cite{EEMR2} with $q_3=0$.

So, as advertised, $a$ and $b$ disappear from the limiting solution.
However, a remnant of their
presence survives in the form of two constraints on the values of
the parameters. Note that since $a^2\geq 0$, the balancing condition
\reef{balanceextr} gives rise to the bound 
\beq
Q_3 \ge q_1 q_2\,,
\label{Q3q1q2}\eeq 
which was also found in
\cite{EEMR2}. Further, 
using that $0\leq b^2\leq a^2$, the Dirac-Misner condition \reef{DMextr} gives
\beq
  R^2 \le \frac{Q_1 Q_2 (Q_3 - q_1 q_2)}{q_1 q_2 (Q_1 + Q_2)}\,,
  \label{CCCbound}
\eeq
which is more meaningfully rewritten,
using \reef{q1q2}, as a bound on a combination of the angular momenta,
\beq
 \frac{4G_5}{\pi}\left(J_\psi-J_\phi\right) \leq \sqrt{\frac{Q_1 Q_2}{q_1 q_2}}
  \left(Q_3 - q_1 q_2\right) \,. 
  \label{DMextr2}
\eeq
The bound \reef{CCCbound} and the constraint \reef{q1q2} are precisely the conditions
found in \cite{EEMR2} in order to avoid closed causal curves for the BPS
solution. It is curious that the condition for eliminating Dirac-Misner
strings in the non-supersymmetric geometry becomes precisely the same as
the condition of avoiding causal pathologies in the supersymmetric
solution. 

\bigskip

The supersymmetric limit of a black ring with two charges $Q_1, Q_2$ and
one dipole $q_3$ does not arise as a special case of the above limit. It
must be taken in a different manner, which we present in sec.~\ref{2ch}.


\setcounter{equation}{0}
\section{Non-supersymmetric black rings in minimal 5D
supergravity}
\label{sec:minimal}

A particular case of interest of our solutions is obtained when
the three charges and the three dipoles are set equal,
\beq
\alpha_1=\alpha_2=\alpha_3\equiv\alpha\,,\qquad \mu_1=\mu_2=\mu_3\equiv\mu\,.
\eeq
Then the three gauge fields in \reef{threepot} associated to each of the
brane components are equal, $A^1=A^2=A^3$, and
the moduli associated to
the size of the dimensions $z_i$ are constant. The solution then becomes
a non-supersymmetric black ring of the minimal supergravity theory in
five dimensions. The action for the bosonic sector of this theory is
\beq
I=-\frac{1}{16\pi G_5}\int \sqrt{-g}\bigg( R-\frac{1}{4}F^2- \frac{1}{6\sqrt{3}}\;
\epsilon^{\mu\alpha\beta\gamma\delta}A_\mu
F_{\alpha\beta}F_{\gamma\delta} \bigg)\,,
\eeq
where $F=dA\equiv \sqrt{3}\;dA^i$.
The form of the solution is obtained in a straightforward manner from
the one in Sec.~\ref{subsec:3chsol}, but since it becomes quite
simpler it is worth giving explicit expressions.
The metric is
\beqa
  \label{min5dmetric}
  ds_\rom{5D}^2 &=& -\frac{1}{h^2_\alpha(x,y)}
\frac{H(x)}{H(y)}
\frac{F(y)}{F(x)}
\Big(dt+\omega_\psi(y)d\psi
+\omega_\phi(x)d\phi\Big)^2\nonumber\\[3mm]
&\;&
+h_\alpha(x,y) F(x) H(x) H(y)^2
\\[3mm]\nonumber
&&\hspace{1cm}
\times
\frac{R^2}{(x-y)^2}\Bigg[
     - \frac{G(y)}{F(y) H(y)^3} \, d\psi^2
     - \frac{dy^2}{G(y)}
     + \frac{dx^2}{G(x)}
     + \frac{G(x)}{F(x) H(x)^3} \, d\phi^2
   \Bigg] \, ,
\eeqa
with $H(\xi)=1-\mu\xi$, and $F$ and $G$ as in \reef{FGH}. The functions
$h_i$ now simplify to
\beq
h_\alpha(x,y)=1+\frac{(\lambda + \mu)(x-y)}{F(x) H(y)}
  \, \sinh^2{\alpha}\,.
\eeq
The one-form $\omega$ has components
\beq
\omega_\psi(y) = R(1+y)\cosh\alpha
  \bigg[
    \frac{C_\lambda}{F(y)}  \cosh^2{\alpha}
   - \frac{3C_\mu}{H(y)}  \sinh^2{\alpha}
  \bigg]
\eeq
and
\beq
  \omega_\phi(x) = - R\;\frac{1-x^2}{F(x)H(x)}\;
    \frac{\lambda+\mu}{1+\lambda}\; C_\lambda
    \sinh^3{\alpha}\,,
\label{wphimin}\eeq
where $C_\lambda$ and $C_\mu\equiv C_i$ are given in \reef{Cs}.
To obtain \reef{wphimin} we have used the condition
\beq
\frac{C_\lambda}{1+\lambda}\sinh^2\alpha=\frac{3C_\mu}{1-
\mu}\cosh^2\alpha\,
\label{DMmin}\eeq
necessary to guarantee the absence of Dirac-Misner strings.

The physical parameters of the solution are
\beqa
M&=&\frac{3\pi R^2}{4G_5}\frac{(\lambda+\mu)(1+\mu)^2}{1-\nu}\cosh 2\alpha\,,\\[3mm]
J_\psi\!\!&=&\!\!\frac{\pi R^3}{2G_5}
\frac{(1-\lambda)^{3/2}(1+\mu)^{9/2}}{(1-\nu)^2}\cosh\alpha
\bigg[\;
  \frac{C_{\lambda}}{1-\lambda}
  \cosh^2 \alpha
-\frac{3C_\mu}{1+\mu}\sinh^2 \alpha
\;\bigg] \;,\\[3mm]
J_{\phi}\!\!&=&\!\!-\frac{3\pi R^3}{G_5}
\frac{\sqrt{1-\lambda}\;(1+\mu)^{7/2}(\lambda+\mu)}{(1-\nu)^2(1-\mu)}
  C_\mu
  \cosh^2\alpha\sinh \alpha\;,\\[3mm]
\mathcal{A}_H &=& 8 \pi^2 R^3 \,
  \frac{(1-\lambda)(\lambda - \nu)^{1/2}(1+\mu)^3 (\nu+ \mu)^{3/2} }{(1-\nu)^2(1+\nu)}
 \Bigg|
  \frac{C_\lambda}{\lambda-\nu} \cosh^2{\alpha}
  +\frac{3C_\mu}{\nu+\mu} \sinh^2{\alpha}
  \Bigg|\cosh{\alpha}  \, .\nonumber\\[1mm]\\[1mm]
T_H^{-1}\!\!&=&\!\!4\pi R\frac{\sqrt{\lambda-\nu}(\mu+\nu)^{3/2}}{\nu(1+\nu)}
\cosh{\alpha} \Bigg|
  \frac{C_\lambda}{\lambda-\nu} \cosh^2{\alpha}
  +\frac{3C_\mu}{\nu+\mu} \sinh^2{\alpha}
  \Bigg|\,.\\[3mm]
\Omega_\psi^{-1}\!\!&=&\!\!R\sqrt{1-\lambda}(1+\mu)^{3/2} \cosh{\alpha} \Bigg|
  \frac{C_\lambda}{\lambda-\nu} \cosh^2{\alpha}
  +\frac{3C_\mu}{\nu+\mu} \sinh^2{\alpha}
  \Bigg|\,.
\eeqa
We have used the Dirac-Misner condition \reef{DMcond} to simplify only
the expression for $J_\phi$.
The charge is obtained from the relation
\beq
Q= \frac{4G_5}{3\pi}M  \tanh 2\alpha\,,
\eeq
and the dipole charge is
\beqa
q&=&-4R\frac{\sqrt{1-\lambda}(1+\mu)^{3/2}}{(1-\nu)(1-\mu)}C_\mu\cosh\alpha\nonumber\\
&=&\frac{8G_5}{3\pi}\frac{J_\phi}{Q}\;.
\label{qmin}\eeqa

The solution contains five parameters, $(\lambda,\nu, \mu, \alpha, R)$,
but the two constraints from absence of Dirac-Misner strings and of
conical defects leave only three independent parameters. This implies
that, out of the four conserved charges of the solution, $(M, Q, J_\psi,
J_\phi)$, at most only three of them, $(M, Q, J_\psi)$, are independent.
Eq.~\reef{qmin} shows that the dipole charge $q$ is not an independent
parameter and therefore there cannot be any continuous violation of
uniqueness. This is in contrast with the situation when the net charge
$Q$ is zero, in which the dipole $q$ is an independent parameter and so
uniqueness is violated by a continuous parameter \cite{RE}. In the
solutions in this paper, the addition of a charge, however small,
implies that the net charge and the dipole charge must be related so as
to avoid Dirac-Misner strings.

One can also argue that the solutions in this section do not exhibit
discrete non-uniqueness, \ie that there are no two (or more) ring
solutions which have the same four conserved charges\footnote{There can
be two solutions with the same values of the independent parameters, say
$(M, Q, J_\psi)$, but then they will be distinguished by the value of
$J_\phi$.}. To see this, fix the scale in the solutions by fixing the
mass. Then define, like in \cite{EEMR2}, dimensionless quantities
$j_{\psi,\phi}\propto J_{\psi,\phi}/\sqrt{G_5 M^3}$, which characterize
the spins for fixed mass, and the (relative) energy above supersymmetry
$m=[M-(3\pi/4G_5)Q]/M$. Note that $m$ depends on $\alpha$ only. For
fixed $m$, impose the balancing condition and the Dirac-Misner
condition. Then there is only one free parameter, say $\mu$, so one can
use this to plot a curve in the ($j_\psi,j_\phi$)-plane showing which
values of $j_\psi$ and $j_\phi$ are allowed. If this curve manages
somehow to self-intersect then we would have two solutions with the same
$(m,j_\psi,j_\phi)$, \ie discrete non-uniqueness. We have checked this
for a large representative set of values of $m$, and found that the
curve does not self-intersect, so uniqueness appears to hold among the
rings in this section. Note, though, that we can expect charged
spherical black holes of minimal supergravity to exist with the same
conserved charges as some of these rings.

The solutions in this section do not admit any non-trivial
supersymmetric limit to BPS rings. A natural conjecture is the existence
of a five-parameter non-BPS ring solution, characterized by $(M, Q,
J_\psi, J_\phi, q)$. This family would allow to describe thermal
excitations above the BPS solution of \cite{EEMR1}, and presumably would
exhibit continuous non-uniqueness, as in \cite{RE}, through the
parameter $q$. One would also expect discrete two-fold non-uniqueness
for fixed parameters $(M, Q, J_\psi, J_\phi, q)$, at least for small
enough values of $Q$ and $J_\phi$, by continuity to the solutions with
$Q=0=J_\phi$, which are known to present this feature \cite{RE}.


\setcounter{equation}{0}
\section{D1-D5-P black rings}
\label{sec:d1d5p}

Solutions where the three charges are interpreted as D1-D5-P charges are
of particular relevance since, near supersymmetry, they will admit a
dual description in terms of a rather well-understood 1+1 supersymmetric
conformal field theory. Here we analyze the most general such solution
obtainable with our methods and also two other important particular
cases obtained by setting some dipoles or charges to zero. We denote by
Q/q a solution with net charges Q and dipole charges q.

\subsection{D1-D5-P/d1-d5-kkm black ring}

The non-supersymmetric black ring with three
charges, D1, D5 and momentum P, and dipole charges d1, d5, and
Kaluza-Klein monopole kkm has metric in the string frame
\beqa
  \label{gen10dmetric}
  ds_\rom{IIB}^2 &=& d\tilde{s}^2_\rom{5D} + \sqrt{\frac{h_2}{h_1}}
  \sqrt{\frac{H_1(y) H_2(x)}{H_1(x) H_2(y)}}
  \, d\mathbf{z}_{(4)}^2
  +\frac{h_3}{\sqrt{h_1 h_2}}
  \frac{H_3(x)}{H_3(y)} \sqrt{\frac{H_1(y) H_2(y)}{H_1(x) H_2(x)}}\,
  \bigg[ dz + A^3 \bigg]^2 \, ,\hspace{4mm}~
\eeqa
where the non-vanishing components of the one-form $A^3$ are given in
\reef{A3t}-\reef{A3phi}, and
\beqa
  d\tilde{s}_\rom{5D}^2 &=& -\frac{1}{h_3 \, \sqrt{h_1 h_2}}
   \frac{F(y)}{F(x)} \sqrt{\frac{H_1(x) H_2(x)}{H_1(y) H_2(y)}} \,
   \bigg[
     dt + \omega_\psi(y) \, d\psi + \omega_\phi(x) \, d\phi
   \bigg]^2 \nonumber \\[2mm]
   && ~~
   + \sqrt{h_1 h_2} F(x) H_3(y)
   \sqrt{H_1(x) H_1(y) H_2(x) H_2(y)}\\[2mm]
    \nonumber
   && ~~~~ \times \frac{R^2}{(x-y)^2}
   \bigg[
     - \frac{G(y)}{F(y) H(y)^3} \, d\psi^2
     - \frac{dy^2}{G(y)}
     + \frac{dx^2}{G(x)}
     + \frac{G(x)}{F(x) H(x)^3} \, d\phi^2
   \bigg] \, .
\eeqa
The dilaton is
\beq
  e^{2\Phi} = \frac{h_2}{h_1} \,
  \frac{H_1(y) H_2(x)}{H_1(x) H_2(y)} \, ,
\eeq
and the components of the RR 2-form potential $C^{(2)}$ are given in
appendix \ref{appB}.

\subsubsection*{6D structure}
\itshape{KK dipole quantization}\normalfont

\bigskip
The solution \reef{gen10dmetric} has a non-trivial structure along the
sixth direction $z$ due to the presence of the term $dz+A^3$. The
quantization of the KK dipole can then be obtained easily by requiring
regularity of the fibration. In order to eliminate the
Dirac string singularity of $A^3$ at $x=+1$ we have to perform a
coordinate (gauge) transformation $z\to\hat z-A^3_\phi(x=+1)$. The
$y$-dependence here cancels if we impose the condition \reef{DMcond},
and then the transformation is,
 \beqa
z\rightarrow \hat
z=z-\frac{2R}{s_3}\left[
 \frac{C_2}{1-\mu_2} {s_2} {c_1}
 +\frac{C_1}{1-\mu_1} {c_2}  {s_1}\right]\phi\;.
            \label{zhat}
 \eeqa
With this, the geometry is free of Dirac singularities at $x=+1$.
However, since $z$ parametrizes a compact
Kaluza-Klein direction, $z\sim z+2\pi R_z$ the coordinate transformation
\reef{zhat} is globally
well-defined only if
 \beqa
 2\pi R_z=\pm\frac{2R}{n_{\mathrm{KK}}s_3}\left[
 \frac{C_2}{1-\mu_2} {s_2} {c_1}
 +\frac{C_1}{1-\mu_1} {c_2}  {s_1}\right]\Delta\phi
\eeqa for some (positive) integer $n_{\mathrm{KK}}$. This condition gives
rise to the KK monopole charge quantization: 
\beqa
q_{\mathrm{KK}}=\mp n_{\mathrm{KK}} R_z=
            -\frac{2R}{s_3}\left[
 \frac{C_2}{1-\mu_2} {s_2} {c_1}
 +\frac{C_1}{1-\mu_1} {c_2}  {s_1}\right]\frac{\Delta\phi}{2\pi}\;,
            \label{qkkquant}
\eeqa
and then $q_{\mathrm{KK}}=q_3$ as given in \reef{q3}.

\begin{flushleft}\itshape{Horizon geometry}\normalfont
\end{flushleft}

Making use of the change of coordinates given by \reef{horcoord},
it is clear that the five-dimensional part of the metric of the
type IIB solution is regular at $y_h=-1/\nu$. Note that
the conformal factors multiplying the different terms in
\reef{gen10dmetric} remain finite and non-zero at
$y_h$. In order to make the term $dz+A^3$
also regular at this point, we perform a change of coordinates 
\beqa
dz=dz'+R(1+y)\Bigg[\frac{C_{\lambda}}{F(y)}c_1c_2s_3
                    -\frac{C_1}{H_1(y)}c_1s_2c_3
                    -\frac{C_2}{H_2(y)}s_1c_2c_3
                    -\frac{C_3}{H_3(y)}s_1s_2s_3\Bigg]
                    \frac{\sqrt{|F(y)|H(y)^3}}{G(y)}dy\,.
                    \nonumber\\
\eeqa 
In the $(v,\psi')$ coordinates
this term becomes manifestly analytic at $y_h$,
\beqa 
dz+A^3=dz'+A^3_tdv+A^3_{\psi}d\psi'+A^3_{\phi}d\phi \,, 
\label{zs}
\eeqa
and imposing the charge quantization condition \reef{qkkquant}, it is
perfectly regular on the horizon.

The horizon in six dimensions is a $U(1)$ fibration over the $S^1\times
S^2$ geometry of the five-dimensional horizon \reef{hormet}. In the
supersymmetric case \cite{EEMR2}, it was found that for certain values
of the parameters the $U(1)$ would Hopf-fiber over the $S^2$ to yield an
$S^3$. In the present case, to find the same result we would have to perform
the coordinate transformation 
\beqa
dz''=dz'+A^3_{\psi}(x,
y_{h})\left(d\psi'+\frac{\omega_\phi(x)}{\omega_\psi(y_h)}d\phi\right) 
\eeqa 
to eliminate the leg along the $S^1$ in the fiber in \reef{zs}, but
given the $x$-dependence this change is not compatible with the global
periodicities of $z$ and $\phi$. Hence, the six-dimensional horizon of
these rings is never globally of the form $S^1\times S^3$. It may still
be, though, that the more general non-supersymmetric solutions that we
have conjectured to exist can actually have such a horizon geometry.

\subsection{D1-D5/kkm black rings and two-charge supertubes}
\label{2ch}

This is the simplest case of a two-charge black ring, with one dipole
charge, that can be connected to a well-understood supersymmetric
configuration in string theory. We will show that the supersymmetric
limit of these black rings results into a two-charge supertube.

A particular case of these solutions (with all $\mu_i=0$) was
constructed in \cite{EE}, but the supersymmetric limit of those rings
could only yield supertubes with half the maximum angular momentum (for
given charges and dipole)\footnote{When comparing to the solutions in
\cite{EE}, the reader should be aware that a slightly different form of
the seed is used, so the functions $F$, $G$ and
parameters $\lambda,\nu$ used below are not the same as those in
\cite{EE}.}. Below we show how the inclusion of dipole parameters
$\mu_i$ allows us to recover supertubes within the full range of allowed
angular momenta. Hence we expect that the present solutions are
sufficient to consistently describe the thermal excitations that keep
$J_\phi=0$ of supertubes with two charges, $Q_1$ and $Q_2$, and angular
momentum $J_\psi$.

\subsubsection{Solution}

In the general solution set $\alpha_3=0$
and $\mu_1 = \mu_2 = 0$.
The string-frame metric for the D1-D5/kkm black ring is
\beqa
  \labell{10dmetric}
  ds^2 &=& ds^2_\rom{5D}
  + \sqrt{\frac{h_2}{h_1}}\,
  \bigg( dz_1^2 + dz_2^2 + dz_3^2 + dz_4^2 \bigg)
  \nonumber\\[2mm] 
  &&
  +\frac{1}{\sqrt{h_1 h_2}}
  \frac{H_3(x)}{H_3(y)}
  \left[
   dz
   -R(1+x)
   \left(\frac{C_\lambda}{F(x)}  {s_1}  {s_2}
   - \frac{C_3}{H_3(x)}  {c_1}  {c_2}
   \right)\, d\phi
  \right]^2  \, ,
\eeqa
where
\beqa
  ds_\rom{5D}^2 &=& -\frac{1}{\sqrt{h_1 h_2}}
   \frac{F(y)}{F(x)} \,
   \left[
     dt + R(1+y)
     \left(
    \frac{C_\lambda}{F(y)}  {c_1}  {c_2}
   -\frac{C_3}{H_3(y)}  {s_1}  {s_2}
  \right) d\psi
   \right]^2 \nonumber\\[2mm]
   &&
   + \sqrt{h_1 h_2} \, F(x) H_3(y)\,
   \frac{R^2}{(x-y)^2}
   \Bigg[
     - \frac{G(y)}{F(y) H_3(y)} \, d\psi^2
     - \frac{dy^2}{G(y)}
     + \frac{dx^2}{G(x)}
     + \frac{G(x)}{F(x) H_3(x)} \, d\phi^2
   \Bigg] \, .
\nonumber\\
\eeqa
Note that now
\beq
  h_i = 1+\frac{(\lambda + \mu_3)(x-y)}{F(x) H_3(y)}
  \, {{s_i}}^2
\eeq
for $i=1,2$. The dilaton is
\beq
  e^{2\Phi} = \frac{h_2}{h_1} \, ,
\eeq
and the non-vanishing components of the RR two-form potential are
\beqa
  C^{(2)}_{tz}
  &=& - \frac{(x-y)}{F(x) H_3(y) \, h_2}(\lambda+\mu_3) c_2 s_2  \, ,\\
  C^{(2)}_{\psi z}
  &=& \frac{R(1+y)}{h_2} \frac{H_3(x)}{H_3(y)}
  \bigg[
   \frac{C_\lambda}{F(x)}  {c_1}  {s_2}
   -\frac{C_3}{H_3(x)}  {s_1}  {c_2}
  \bigg]  \, ,\\[2mm]
  C^{(2)}_{t \phi}
  &=&
  -\frac{R(1+x)}{h_2} \frac{F(y)}{F(x)}
  \bigg[
    \frac{C_\lambda}{F(y)}  {s_1}  {c_2}
    -\frac{C_3}{H_3(y)}  {c_1}  {s_2}
  \bigg] \, ,\\[2mm]\nonumber
  C^{(2)}_{\psi \phi}
  &=& -\frac{R^2}{2} \sinh{2\alpha_1} \,
  \Bigg\{
    \frac{G(x)}{(x-y)}
    \frac{\lambda+\mu_3}{F(x) H_3(x)}
    +(1+x)
    \left(
      \frac{C_\lambda^2}{\lambda F(x)}
      +\frac{C_3^2}{\mu_3 H_3(x)}
    \right) \\[2mm]
    &&~~~~
    +\frac{1}{h_2}(1+x)(1+y) \frac{H_3(x)}{H_3(y)}
    \bigg[
      \frac{C_\lambda^2}{F(x)^2}  {s^2_2}
        +\frac{C_3^2}{H_3(x)^2}  {c^2_2}
    \bigg]   \Bigg\}  \\[2mm]\nonumber
    &&~~~~
  + \frac{(1+x)(1+y)}{h_2} \frac{R^2}{2}
    \cosh{2\alpha_1} \sinh{2\alpha_2} \,
    \frac{C_\lambda C_3}{F(x) H_3(y)} \, .
\eeqa
Since $\omega_\phi=0$ there are no dangerous Dirac-Misner strings in this solution.

\subsubsection{Properties}
The ADM mass of the black ring is
\beqa
  M = \!\!\frac{\pi R^2}{4G_5} \frac{1}{1-\nu}
\Bigg\{\lambda-\mu_3
       +2 \lambda\mu_3
  + (\lambda+\mu_3)( \cosh{2\alpha_1}+ \cosh{2\alpha_2})
\Bigg\} \, ,
\eeqa
and the angular momentum is
\beq
  J_{\psi} = \frac{\pi R^3}{2G_5}
\frac{\sqrt{(1-\lambda)(1+\mu_3)}}{(1-\nu)^2}
\bigg[\;
  (1+\mu_3)C_{\lambda}
   {c_1}  {c_2}
  -(1-\lambda)C_3
   {s_1}  {s_2}
\;\bigg] \, .
\eeq
The D1- and D5 charges are
\beq
  Q_\rom{D1} = \frac{R^2 (\lambda+\mu_3)\sinh{2\alpha_2}}{1-\nu}\, ,
  ~~~~~~
  Q_\rom{D5} = \frac{R^2 (\lambda+\mu_3)\sinh{2\alpha_1}}{1-\nu}\, ,
\eeq
and the dipole charge from the tubular Kaluza-Klein-monopole
is\footnote{This is the same result as for $q_3$ in \reef{q3} but with
the sign reversed so as to agree with the sign choices 
in \cite{EE,EEMR2}.}
\beq
  q_\rom{KK} = 2 R
   \bigg[ \frac{C_\lambda}{1+\lambda}  {s_1}  {s_2}
   - \frac{C_3}{1-\mu_3}  {c_1}  {c_2}
   \bigg] \frac{\Delta\phi}{2\pi} \, .
\eeq
The horizon area is
\beqa
  \mathcal{A}_H &=& 8 \pi^2 R^3 \,
  \frac{1}{(1-\nu)^2(1+\nu)}
  \bigg|
  \epsilon_\lambda\sqrt{\lambda (1-\lambda^2)(\mu_3+\nu)}(1+\mu_3)
    {c_1} {c_2}  
\nonumber\\ 
   && \hspace{4.3cm}
   +\epsilon_3\sqrt{\mu_3 (1-\mu_3^2)(\lambda-\nu)}(1-\lambda)
    {s_1} {s_2}
  \bigg| \, .
\eeqa

\subsubsection{Supersymmetric limit: Two-charge supertubes}

Now take the supersymmetric limit $\alpha_1\sim\alpha_2 \to \infty$ and $\mu_3\sim
\nu\sim\lambda \to 0$ keeping fixed
\beq
  (\lambda + \mu_3) e^{2\alpha_1} = \frac{2 Q_\rom{D5}}{R^2} \, ,
  \hspace{8mm}
  (\lambda + \mu_3) e^{2\alpha_2} = \frac{2 Q_\rom{D1}}{R^2} \, .
\eeq
Note that the dipole charge remains finite in the limit,
\beq
  q_\rom{KK} = \frac{R}{2}
  \left( \epsilon_\lambda \sqrt{\lambda (\lambda - \nu)}
         -  \epsilon_3 \sqrt{\mu_3 (\mu_3 + \nu) }\right)
   e^{\alpha_1 + \alpha_2} \, .
\eeq
Taking the supersymmetric limit gives the string-frame metric
\beqa
 ds^2 =
  -\frac{1}{\sqrt{h_1 h_2}}
  \left(dt + \frac{q_\rom{KK}}{2}(1+y)d\psi \right)^2
  +\sqrt{h_1 h_2} \, d{\bf x}_4^2
  +\frac{1}{\sqrt{h_1 h_2}}
  \left(dz - \frac{q_\rom{KK}}{2}(1+x) d\phi \right)^2
  +\frac{h_2}{h_1} d{\bf z}_4^2
\nonumber\\
\eeqa
with
\beq
  h_1 = 1 +\frac{Q_\rom{D5}}{2R^2}(x-y) \, , \hspace{8mm}
  h_2  = 1 +\frac{Q_\rom{D1}}{2R^2}(x-y) \, .
\eeq
The rest of the fields in the solution are also easily obtained.

This is
exactly the 2-charge/1-dipole BPS solution of \cite{EEMR2}, which is the
same as the two-charge supergravity supertube of \cite{luma,EMT}.
However, it is important to
realize that if we approach the supersymmetric limit
through a sequence of regular non-supersymmetric black rings, then the
condition that they are balanced imposes restrictions on the
parameters. Specifically, balancing the ring requires, in the limit,
$2\nu = \lambda - \mu_3$.
It is now convenient to define 
\beq
k = \frac{2 \mu_3}{\lambda + \mu_3}\,.
\eeq 
Since we must
have $0 \le \mu_3 \le \lambda < 1$, $k$ takes values between 0 and 1.
Defining the function
\beq
  f(k) = \frac{1}{2}
   \left( \epsilon_\lambda \sqrt{2-k} - \epsilon_3 \sqrt{k}\right)\, ,
  \label{funcf}
\eeq
we can then write
\beq
  q_\rom{KK} = \frac{\sqrt{Q_\rom{D1}Q_\rom{D5}}}{R} f(k) \, .
\label{qkk}\eeq
This can be used to eliminate $R$ (which is not a proper invariant
quantity of the supergravity solutions) from the angular momentum
\beq
  J_\psi = \frac{\pi}{4G_5} R^2 \, q_\rom{KK} \, ,
\eeq
and then express it in terms of
the number of
branes and Kaluza-Klein monopoles as\footnote{See \eg \cite{EE,EEMR2} for the
expressions for the quantized brane numbers in terms of supergravity
charges. Note that we take $n_\mathrm{KK}$ to be positive.}
\beq
  J_\psi = \frac{N_\rom{D1} N_\rom{D5}}{n_\rom{KK}}\,\mathrm{sgn}[f(k)]\, f(k)^2\, .
\eeq
Using the four
combinations of $\epsilon_\lambda , \epsilon_3 = \pm 1$, the values of
the function $f(k)$ cover precisely the interval $[-1,1]$ when $k$ is
varied between $0$ and $1$. Thus the supergravity solution
yields exactly the range of angular momentum
\beq
  |J_\psi| \le \frac{N_\rom{D1} N_\rom{D5}}{n_\rom{KK}} \,
\label{jbound}\eeq
expected from the worldvolume supertube analysis
\cite{MT,EMT,MNT,HO}. The
solutions constructed earlier in \cite{EE}
correspond instead to having $\mu_3=0$, hence $k=0$ and $f^2(0)=1/2$.
Since we can now vary $\mu_3$ and recover the whole range of expected
values of the
parameters, it seems that the solution above should be the most general
non-supersymmetric black ring with two charges and one angular momentum.

The supersymmetric supertube solutions with angular momentum strictly
below the bound \reef{jbound} have naked singularities, while if the
bound is saturated the solution is regular \cite{flow}. The saturation
of the bound as a limit of our solutions is slightly subtle, since one
must take $k=1$ \ie $\mu_3= \lambda$. For finite values of the
parameters this would require $\nu=0$, which corresponds to an extremal
singular solution. So it would seem that in order to thermally deform
the non-singular, maximally rotating supertube, one has to consider
$\mu_3$ strictly different from $\lambda$. This suggests that the energy
gap of non-BPS excitations above the non-singular, maximally rotating
supertube is larger than for under-rotating supertubes. Indeed this is
expected, since in the former all the effective D1-D5 strings are singly
wound (it is a ground state of the non-twisted sector of the dual CFT)
and therefore the left- and right-moving open strings are energetically
more expensive to excite.

Finally, on the issue of non-uniqueness, the solutions contain five parameters
(\ie ($\alpha_{1,2}$, $\mu_3$, $\nu$, $\lambda$, $R$), minus one constraint) but
only four conserved charges $(M,Q_{1,2},J_\psi)$, so we have a
continuous one-parameter non-uniqueness. This is of course controlled
by the dipole charge $q_\mathrm{KK}$. It is possible to see this
explicitly by drawing plots of the area of the rings vs.\ $J_\psi$
at different values of $q_\mathrm{KK}$, for fixed values of the mass and
the charges. These curves look very similar to the ones for the dipole
black ring with $N=1$ in \cite{RE}, so we shall not reproduce them here.


\subsection{D1-D5-P/d1-d5 black ring: the black double helix}

The supersymmetric configuration in the previous subsection is U-dual to
a D1-P/d1 supergravity supertube. This is the supergravity description
of a D1-helix: a D1-brane with a gyrating momentum wave, such that it
coils around the $\psi$ direction (hence the dipole d1) while carrying
momentum along $z$. The supergravity solution is smeared along the
direction of the tube \cite{luma}. T-dualizing along the internal $T^4$
directions one obtains a D5-P/d5 configuration, \ie a D5-helix.

The D1 and D5 can bind to form a supersymmetric double D1-D5 helix. To
obtain a non-supersymmetric black D1-D5 helix (without a kkm tube),
we must first demand
\beq
q_3=0 ~~\mbox{and}~~ q_1, q_2 \neq 0\,.
\label{q3zero}\eeq
However, if we want the configuration to actually describe the
excitations of a bound state of the D1-D5 helix, we must also require 
\beq
q_1Q_1=q_2Q_2\,
\label{qQ}
\eeq
even away from the supersymmetric state. Since $Q_1$ and $q_2$ ($Q_2$
and $q_1$) are associated to the number of windings of D5-branes
(D1-branes) along $z$ and along $\psi$, respectively, this equation is
the condition that the pitches of the D1 and D5 helices are equal (see
also \cite{EEMR2}). The two helices can then bind and these black tubes
are naturally interpreted as thermal excitations of a D1-D5 superhelix.

It is straightforward to see that the two conditions \reef{q3zero} and
\reef{qQ} imply
\beq
\mu_1=\mu_2=0\,,
\label{mu1mu2}
\eeq 
while all other parameters take non-zero values. 

The supersymmetric limit of these solutions, as shown in section
\ref{sec:extsusy}, correctly yields
the BPS solutions with three charges and two dipoles of \cite{EEMR2}.
However, \reef{mu1mu2} implies that the two ``constraint parameters''
$a$ and $b$ in \reef{a2}, \reef{b2}, are not independent but satisfy
$a^2=2b^2$. From \reef{balanceextr} and \reef{DMextr} we obtain again the
inequality \reef{Q3q1q2}, but now, instead of the bound \reef{CCCbound}, we
find that
\beq 
\frac{4G_5}{\pi}\left(J_\psi-J_\phi\right) =
\frac{1}{2}\sqrt{\frac{Q_1 Q_2}{q_1 q_2}} \left(Q_3 - q_1 q_2\right) \,,
\label{booo} 
\eeq
\ie we can only obtain half the maximum value for $J_\psi-J_\phi$. This
restriction is reminiscent of the former situation for D1-D5/kkm
supertubes in \cite{EE}, and suggests that we need a still larger
family of non-BPS solutions in order to describe the thermal
deformations of supertubes with three charges and two dipoles and
generic values of $J_\psi-J_\phi$.

\subsection{Decoupling limit}
\label{sec:decoupling}

The decoupling limit, relevant to AdS$_3$/CFT$_2$ duality of the D1-D5
system, is obtained by taking the string length to zero,
$\alpha'=\ell_s^2\to 0$, and then scaling the parameters in the solution
in such a way that
\beq
\lambda,\quad \nu,\quad \mu_i,\quad \alpha_3,\quad
\alpha'e^{2\alpha_{1,2}},\quad R/\alpha', 
\eeq
remain fixed. The (dimensionless) coordinates $x,y$ also remain finite.
Note in particular that the boosts associated to D5- and D1-brane
charges become $\alpha_{1,2}\to\infty$, and that the energy of the
excitations near the core of the solution is kept finite by scaling the
parameter $R\sim \alpha'$. To keep the ten-dimensional string metric
finite, we rescale it by an overall factor of $\alpha'$. 

The metric in the decoupling limit is of the same form as
\reef{gen10dmetric}, with the functions $F$, $G$, $H_i$ and $h_3$
unmodified, but $h_{1,2}$, $\omega$ and $A^3$ do change. One can see in
general that the metric asymptotes to AdS$_3\times S^3\times T^4$, but
we shall provide details only for the two cases of interest of the
previous subsections, with $\mu_1=\mu_2=0$ (and in general $\alpha_3\neq
0$), where the expressions get somewhat simplified. In this case one
gets
\beq
h_{1,2}=\frac{Q_{1,2}}{2R^2}(1-\nu)\frac{x-y}{F(x)H_3(y)}\,,
\eeq
and
\beqa
\omega_\psi&=&\frac{G_5 J_\psi}{\pi R^2}\frac{2(1+y)}{F(y)H_3(y)}
\frac{C_\lambda H_3(y)-C_3 F(y)}{C_\lambda (1+\mu_3)-C_3 (1-\lambda)}
\frac{(1-\nu)^2}{\sqrt{(1-\lambda)(1+\mu_3)}}\,,\nonumber\\
\omega_\phi&=&\frac{G_5 J_\phi}{\pi R^2}\frac{1-x^2}{F(x)H_3(x)}
\frac{(1-\nu)^2}{\sqrt{(1-\lambda)(1+\mu_3)}}\,,
\eeqa
while the expression for $A^3$ is not comparatively simpler. The
asymptotic region is again at $x\to y\to -1$. If we perform the change
of coordinates
\beq
r^2=R^2 \frac{1-x}{x-y}\frac{(1-\lambda)(1+\mu_3)}{1-\nu}\,,\qquad
\cos^2\theta=\frac{1+x}{x-y}\,,
\eeq
introduce canonical angular variables \reef{canang}, and
gauge-transform so that $A^3_t$
vanishes at infinity, then the asymptotic metric at $r\to\infty$ is 
\beq
ds^2_\mathrm{IIB}\to
\frac{r^2}{\sqrt{Q_1Q_2}}\left(-dt^2+dz^2\right)+\sqrt{Q_1
Q_2}\;\frac{dr^2}{r^2}
+\sqrt{Q_1 Q_2}\left(d\theta^2+\sin^2\theta d\tilde\psi^2+\cos^2\theta
d\tilde\phi^2\right)+\sqrt{\frac{Q_2}{Q_1}}d\mathbf{z}_{(4)}^2\,,
\label{asympads}\eeq
\ie the correct asymptotic geometry of
AdS$_3\times S^3\times T^4$.

Away from the asymptotic boundary the geometry does not factorize into a
product space AdS$_3\times S^3\times T^4$, not even locally. In general,
the geometry does not appear to become simple even near the horizon.
This is in contrast to what was found in the supersymmetric case with
three charges and three dipoles, where near the horizon a second,
different factorization into (locally) AdS$_3\times S^3\times T^4$
happens \cite{EEMR2,BK2}. With three charges and only two dipoles, this
factorization does not happen even near the core of the supersymmetric
solutions. But perhaps the yet to be found non-BPS solution that can
retain all three dipoles in the supersymmetric limit will show this
property.

\setcounter{equation}{0}
\section{Discussion}
\label{sec:discuss}

We have shown how to construct non-supersymmetric three-charge black
rings via boosts and dualities, overcoming the problems previously
encountered in ref.~\cite{EE}. Technically, the main issue is the
requirement that the one-form $\omega$ associated to the rotation of the
ring be well-defined (absence of Dirac-Misner strings). This was also an
important ingredient when constructing the supersymmetric rings by
solving the supersymmetry-preservation equations of \cite{hetal}. In the
present case, we have achieved it by having more parameters, associated
to dipole charges, in our seed solution. 

We expect that a larger family of non-supersymmetric black rings with
nine-parameters $(M,J_\psi,J_\phi,Q_{1,2,3},q_{1,2,3})$ exists, and that
the general solutions of \cite{EEMR1,EEMR2,BW,JGJG2} are recovered in the
supersymmetric limit. The nine parameters would yield three-fold
continuous non-uniqueness, furnished by the non-conserved dipole charges
$q_{1,2,3}$. This is like in the dipole ring solutions in \cite{RE} but
larger than the two-fold continuous non-uniqueness of the supersymmetric
rings of \cite{EEMR2}. Presumably, one of the additional parameters in
the conjectured larger family of solutions will not describe proper
near-supersymmetric excitations of the supertube. Instead they should be
interpreted as the presence of dipole strings or branes in the
configuration that are not bound to the supertube but rather
superimposed on it. Since these dipole branes break all supersymmetries,
they must disappear in the BPS limit.

The extra parameters that are still missing from our current solutions
should provide the freedom to vary the angular momentum $J_\phi$ and all
the three dipoles independently of other parameters. In particular, we
expect that black rings with two charges should exist that carry both
angular momenta $J_\psi$ and $J_\phi$. Note, though, that in the
supersymmetric limit $J_\phi$ disappears, which is consistent with the
fact that $J_\phi$ is expected to be carried by the coherent
polarization of (R-charged) left- and right-moving fermionic open string
excitations. In our solutions the total macroscopic $J_\phi$ must
vanish. So the two-charge solutions in this paper can describe the
non-BPS excitations that carry total $J_\phi=0$.\footnote{We could have
two-charge supertubes with $J_\phi\neq 0$ if $\mu_1,\mu_2\neq 0$. But
then $q_1,q_2\neq 0$, so these excitations take the system further away
from the supersymmetric state.} This should be already enough to address
in detail the issues of black hole non-uniqueness and
near-supersymmetric black ring entropy from a microscopic viewpoint,
along the lines proposed in ref.~\cite{EE}. We hope to tackle these
problems in the future.

The solutions with three charges and two dipoles in this paper are
similarly expected to describe the non-BPS excitations that do not add
to the $J_\phi$ of a D1-D5-P/d1-d5 super-helix. However, in this case we
seem to be restricted to considering only excitations above the
supersymmetric state with half the maximum value of $J_\psi-J_\phi$.
This is analogous to the former situation in \cite{EE}, and once again
points to the need to find a larger family of solutions. Note, however,
that by including more general excitations our solutions do describe
thermally deformed D1-D5-P/d1-d5 supertubes with any value of
$J_\psi-J_\phi$ in the permitted range \reef{DMextr2}. As we have seen,
these solutions are free of pathologies and are regular on and outside
the horizons.

One can speculate about generalizations of the non-supersymmetric black
rings. Refs.~\cite{BW,BWW} constructed supersymmetric ring solutions
with arbitrary cross-sections. However, it was argued in \cite{GHHR}
that unless the cross-section is circular these rings are not truly
black holes, since they do not have smooth horizons. Adding energy to a
ring with a non-circular cross-section is unlikely to yield a stationary
black ring, since the lumpiness of the rotating ring will presumably
cause the system to radiate when it is not supersymmetric. 

Finally, a particularly interesting spin-off of our study is the
evidence we have found in favor of the proposal in ref.~\cite{gubser}
for one criterion to admit naked singularities in supergravity
solutions. The supersymmetric limits of our black rings do indeed
typically result in solutions with naked singularities.
Ref.~\cite{gubser} proposed that solutions which admit thermal
deformations, \ie arise as the zero-temperature limit of black holes
with a regular horizon, should be regarded as physically sensible. This
is precisely what we have found, and in a very non-trivial manner.
Taking the supersymmetric limit of our rings results in supertubes with
parameters precisely within the ranges that were earlier determined
using entirely different criteria --- worldvolume theory constraints
(including unitarity), and absence of localized causal violations in BPS
solutions. Solutions with sick worldvolume theories, or spacetime causal
pathologies, such as over-rotating supertubes, do not admit thermal
deformations. This is a remarkable consistency check of the low energy
supergravity description of string theory configurations.

\section*{Acknowledgments}

\noindent We thank Gary Horowitz, David Mateos, Harvey Reall and Toby
Wiseman for useful discussions. We would also like to thank the
organizers and the participants at the Berkeley mini-conference on black
rings for lively discussions. HE was supported by NSF grant PHY-0244764
and the Danish Research Agency. RE and PF were supported in part by
CICyT grant FPA2001-3598, by DURSI 2001-SGR-00188 and by the European
Comission RTN program under contract MRTN-CT-2004-005104. RE was
additionally supported in part by UPV00172.310-14497, and PF by a FI
scholarship from Generalitat de Catalunya.

\appendix

\section*{Appendices}

\setcounter{equation}{0}
\section{Three-form fields for the 11D solution}
\label{appA}

We here give explicitly the non-zero components of the three-form
potential for the eleven-dimensional solution in section \ref{sec:11dsoln}:
\beqa
A^1_t &=&\frac{U_1-1}{h_1}  c_1 s_1 \;,
\\
A^1_\psi &=&\frac{R(1+y)}{h_1}
\Bigg[
  U_1\frac{C_{\lambda}}{F(y)} s_1  c_2  c_3
  -U_1\frac{C_1}{H_1(y)} s_1  s_2  s_3
  -\frac{C_2}{H_2(y)} c_1  c_2  s_3
  -\frac{C_3}{H_3(y)} {c_1}  s_2  c_3
\Bigg], \\
A^1_\phi &=&-\frac{R(1+x)}{h_1}
\Bigg[
  \frac{C_{\lambda}}{F(x)} c_1  s_2  s_3
   -\frac{C_1}{H_1(x)} c_1  c_2  c_3
   -U_1\frac{C_2}{H_2(x)} s_1  s_2  c_3
   -U_1\frac{C_3}{H_3(x)} s_1  c_2  s_3
\Bigg] , \\
A^2_t &=&\frac{U_2-1}{h_2} c_2 s_2 \;,
\\
A^2_\psi &=&\frac{R(1+y)}{h_2}
\Bigg[
  U_2\frac{C_{\lambda}}{F(y)} c_1  s_2  c_3
  -\frac{C_1}{H_1(y)} {c_1}  {c_2}  {s_3}
  -U_2\frac{C_2}{H_2(y)} {s_1}  {s_2}  {s_3}
  -\frac{C_3}{H_3(y)} s_1  c_2  c_3
\Bigg] , \\
A^2_\phi &=&-\frac{R(1+x)}{h_2}
\Bigg[
  \frac{C_{\lambda}}{F(x)} s_1  c_2  s_3
  -U_2\frac{C_1}{H_1(x)} s_1  s_2  c_3
  -\frac{C_2}{H_2(x)} c_1  c_2  c_3
  -U_2\frac{C_3}{H_3(x)} {c_1}  s_2  s_3
\Bigg] , \\
A^3_t &=&\frac{U_3-1}{h_3} c_3 s_3 \;, \label{A3t}
\\
A^3_\psi &=&\frac{R(1+y)}{h_3}
\Bigg[
  U_3\frac{C_{\lambda}}{F(y)} {c_1}  {c_2}  {s_3}
  -\frac{C_1}{H_1(y)} {c_1}  {s_2}  {c_3}
  -\frac{C_2}{H_2(y)} {s_1}  {c_2}  {c_3}
  -U_3\frac{C_3}{H_3(y)} {s_1}  {s_2}  {s_3}
\Bigg], \\
\label{A3phi}
A^3_\phi &=&-\frac{R(1+x)}{h_3}
\Bigg[
  \frac{C_{\lambda}}{F(x)} {s_1}  {s_2}  {c_3}
  -U_3\frac{C_1}{H_1(x)} {s_1}  {c_2}  {s_3}
  -U_3\frac{C_2}{H_2(x)} {c_1}  {s_2}  {s_3}
  -\frac{C_3}{H_3(x)} {c_1}  {c_2}  {c_3}
\Bigg].
\eeqa
where the functions $U_i$ are defined as
\beq
  U_i = \frac{F(y) H(x)^3}{F(x) H(y)^3}
 \frac{H_i(y)^2}{H_i(x)^2} \, .
\label{Ui}\eeq


\setcounter{equation}{0}
\section{The RR two-form potentials for the D1-D5-P black ring
solution}
\label{appB}
We here give the expressions for the non-zero components of the
Ramond-Ramond two-form potential of the D1-D5-P non-supersymmetric black
ring solution of type IIB supergravity given in section \ref{sec:d1d5p}:
\beqa
  C^{(2)}_{tz}
  &=& \frac{U_2-1}{h_2} c_2 s_2  \, ,\\
  C^{(2)}_{\psi z}
  &=& \frac{R(1+y)}{h_2}
  \bigg[
    U_2 \, \frac{C_\lambda}{F(y)}  {c_1}  {s_2}  {c_3}
   - \frac{C_1}{H_1(y)}  {c_1}  {c_2}  {s_3} 
   - U_2 \, \frac{C_2}{H_2(y)}  {s_1}  {s_2}  {s_3}
   -\frac{C_3}{H_3(y)}  {s_1}  {c_2}  {c_3}
  \bigg]  ,\\[2mm]
  C^{(2)}_{\phi z}
  &=& - \frac{R(1+x)}{h_2}
  \bigg[
    \frac{C_\lambda}{F(x)}  {s_1}  {c_2}  {s_3}
   - U_2  \frac{C_1}{H_1(x)}  {s_1}  {s_2}  {c_3} 
   - \frac{C_2}{H_2(x)}  {c_1}  {c_2}  {c_3}
   -U_2 \frac{C_3}{H_3(x)}  {c_1}  {s_2}  {s_3}
  \bigg], \\[2mm]
  C^{(2)}_{t \psi}
  &=&
  \frac{R(1+y)}{h_2}
  \bigg[
    U_2 \, \frac{C_\lambda}{F(y)}  {c_1}  {s_2}  {s_3}
   - \frac{C_1}{H_1(y)}  {c_1}  {c_2}  {c_3} 
   - U_2 \, \frac{C_2}{H_2(y)}  {s_1}  {s_2}  {c_3}
   -\frac{C_3}{H_3(y)}  {s_1}  {c_2}  {s_3}
  \bigg],
  \\[2mm]
  C^{(2)}_{t \phi}
  &=&
  -\frac{R(1+x)}{h_2}
  \bigg[
    \frac{C_\lambda}{F(x)}  {s_1}  {c_2}  {c_3}
   - U_2 \frac{C_1}{H_1(x)}  {s_1}  {s_2}  {s_3} 
   - \frac{C_2}{H_2(x)}  {c_1}  {c_2}  {s_3}
   -U_2  \frac{C_3}{H_3(x)}  {c_1}  {s_2}  {c_3}
  \bigg],\\[2mm]
  C^{(2)}_{\psi \phi}
  &=& -R^2 c_1 s_1 \,
  \Bigg\{
    \frac{G(x)}{(x-y)}
    \left(
      \frac{\lambda+\mu_3}{F(x) H_3(x)}
      +\frac{\mu_2-\mu_1}{H_1(x) H_2(x)}
    \right) \nonumber\\
    &&~~~~
    +\frac{1}{h_2}(1+x)(1+y)
    \bigg[
      U_2\, \frac{C_\lambda^2}{F(x)F(y)}  {s^2_2}
      -\frac{C_1^2}{H_1(x)H_1(y)}  {c^2_2} \nonumber\\
     &&\hspace{4.5cm}
      -U_2\, \frac{C_2^2}{H_2(x)H_2(y)}  {s^2_2}
      +\frac{C_3^2}{H_3(x)H_3(y)}  {c^2_2}
    \bigg] \nonumber\\
    &&~~~~
    +(1+x)
    \left(
      \frac{C_\lambda^2}{\lambda F(x)}
      -\frac{C_1^2}{\mu_1 H_1(x)}
      +\frac{C_2^2}{\mu_2 H_2(x)}
      +\frac{C_3^2}{\mu_3 H_3(x)}
    \right)
  \Bigg\} \nonumber\\[2mm]
  &&~~
  + \frac{(1+x)(1+y)}{h_2} R^2 c_2 s_2
   \Bigg\{
     \frac{C_\lambda C_3}{F(x) H_3(y)}  {s^2_1}
     + U_2\, \frac{C_\lambda C_3}{F(y) H_3(x)}  {c^2_1} \nonumber\\
    &&\hspace{4cm}
     +U_2\frac{C_1 C_2}{H_1(x) H_2(y)}  {s^2_1}
     + \frac{C_1 C_2}{H_1(y) H_2(x)}  {c^2_1}
   \Bigg\} \, ,
\eeqa
where $U_i$ were defined in \reef{Ui}.

\setcounter{equation}{0}
\section{Limit of spherical black hole}
\label{appC}

We can take a limit of our solutions to recover non-supersymmetric
spherical black holes with three charges. This limit is actually the
same as described in \cite{RE},
and involves taking $\lambda,\nu\to 1$ and $R\to 0$ while keeping finite
the parameters $a$, $m$, defined as
\beq
m=\frac{2R^2}{1-\nu}\,,\qquad a^2=2R^2\frac{\lambda-\nu}{(1-\nu)^2}\,.
\eeq
The coordinates
$x$, $y$ degenerate in this limit, so we introduce new ones,
$r$, $\theta$, through
\beqa
x&=&-1+2\left(1-\frac{a^2}{m}\right)\frac{R^2\cos^2\theta}{r^2-(m-
a^2)\cos^2\theta}\,,\nonumber\\
y&=&-1-2\left(1-\frac{a^2}{m}\right)\frac{R^2\sin^2\theta}{r^2-(m-
a^2)\cos^2\theta}\,,
\eeqa
and rescale $\psi$ and $\phi$
\beq
(\psi,\phi)\to \sqrt{\frac{m-a^2}{2R^2}}\;(\psi,\phi)
\eeq
so they have canonical periodicity $2\pi$. Then we recover the metric
\beqa
ds^2_\rom{5D}&=&-(h_1h_2h_3)^{-2/3}\left(1-\frac{m}{\Sigma}\right)
\left(dt-\frac{m a\sin^2\theta}{\Sigma-m}\: c_1 c_2 c_3 \:d\psi
-\frac{m a\cos^2\theta}{\Sigma}\: s_1 s_2 s_3 \:d\phi
\right)^2\nonumber\\
&&+(h_1h_2h_3)^{1/3}\left[\Sigma\left(\frac{dr^2}{\Delta}+d\theta^2\right)
+\frac{\Delta\sin^2\theta}{1-
m/\Sigma}\:d\psi^2+r^2\cos^2\theta\:d\phi^2\right]\,,
\eeqa
\beq
\Delta\equiv r^2-m+a^2\,,\qquad \Sigma\equiv
r^2+a^2\cos^2\theta\,,\qquad h_i=1+\frac{m s_i^2}{\Sigma}\,.
\eeq
This is the particular case of the rotating black hole with three
charges in \cite{BLMPSV} that is obtained by setting one of the two
rotation parameters of the initial seed black hole to zero.

Note that we have not prescribed any limiting value for the parameters
$\mu_i$. As was the case for the dipole rings in \cite{RE}, in the limit
all the functions $H_i(\xi)\to 1+\mu_i$ become constants that can be
absorbed in rescalings of the coordinates. Then the limiting
spherical black hole solution is actually independent of these
parameters and therefore they cannot provide it
with any kind of `hair'.

\setcounter{equation}{0}
\section{Infinite radius limit}
\label{appD}

In the limit where the radius of the $S^1$ of the ring becomes infinite
the ring becomes a black string carrying momentum along its length. The
dipole charges become conserved charges, so in this limit we obtain a
five-dimensional black string with six charges and momentum. Reduction
to four dimensions along the length of the string results in a
non-supersymmetric four-dimensional black hole with seven charges. Their
extremal limit in this case is also a supersymmetric limit.

We take $R \to \infty$ while keeping fixed
\beq
  r = -\frac{R}{y} \, ,~~~~~~
  \cos{\theta} = x \, ,~~~~~~
  \eta = R\psi \, .
\eeq
In order to get a finite limit, we also take $\lambda, \nu, \mu_i \to 0$
keeping fixed
\beq
  r_0 = \nu R \, ,~~~~~~
  r_0 \cosh^2{\sigma}= \lambda R \, ,~~~~~~
  r_0 \sinh^2{\gamma_i}= \mu_i R \, .
\eeq
Note first that the balancing condition gives
\beq
  \sinh^2{\sigma} = 1 + \sum_{i=1}^3 \sinh^2{\gamma_i}
\eeq
and the Dirac-Misner condition \reef{DMcond} becomes
\beq
  0 = \omega_\phi
  \; = \;
  \epsilon_\lambda
    \sinh{2\sigma} {s_1} {s_2} {s_3}
    -\epsilon_1
    \sinh{2\gamma_1} {s_1} {c_2} {c_3}
   -\epsilon_2
    \sinh{2\gamma_2} {c_1} {s_2} {c_3}
   -\epsilon_3
    \sinh{2\gamma_3} {c_1} {c_2} {s_3}
  \, .
\eeq
The metric becomes
\beqa
  ds_\rom{11D}^2 &=&
  -  \frac{1}{(h_1 h_2 h_3)^{2/3}}
  \frac{\hat{f}}{(\hat{h}_1 \hat{h}_2 \hat{h}_3)^{1/3}}
  \Big[ dt + \omega_\eta \, d\eta \Big]^2 \\[2mm]
  \nonumber &&\hspace{5mm}
  +(h_1 h_2 h_3)^{1/3} \Bigg\{
  \frac{f}{\hat{f} (\hat{h}_1 \hat{h}_2 \hat{h}_3)^{1/3}}d\eta^2
  +(\hat{h}_1 \hat{h}_2 \hat{h}_3)^{2/3}
  \left( \frac{dr^2}{f} + r^2 d\Omega_2^2
  \right) \\[2mm] \nonumber
  &&\hspace{1.5cm}
  \frac{1}{(\hat{h}_1 \hat{h}_2 \hat{h}_3)^{1/3}}
  \left[
    \frac{\hat{h}_1}{h_1} (dz_1^2 + dz_2^2)
    +\frac{\hat{h}_2}{h_2} (dz_3^2 + dz_4^2)
    +\frac{\hat{h}_3}{h_3} (dz_5^2 + dz_6^2)
  \right]\Bigg\}  \, ,
\eeqa
where
\beq
  f = 1 - \frac{r_0}{r}\, ,~~~~~~
  \hat{f} = 1 - \frac{r_0 \cosh^2{\sigma}}{r}\, ,~~~~~~
  \hat{h}_i = 1 + \frac{r_0 \sinh^2{\gamma_i}}{r}\, ,
\eeq
and
\beqa
  \omega_\eta &=& - \frac{r_0}{2 r}
  \bigg[
    \hat{f}^{-1}\epsilon_\lambda
    \sinh{2\sigma} {c_1} {c_2} {c_3}
    -\hat{h}_1^{-1}\epsilon_1
    \sinh{2\gamma_1} {c_1} {s_2} {s_3}
 \nonumber\\[2mm]
  && \hspace{1cm}
   -\hat{h}_2^{-1}\epsilon_2
    \sinh{2\gamma_2} {s_1} {c_2} {s_3}
   -\hat{h}_3^{-1}\epsilon_3
    \sinh{2\gamma_3} {s_1} {s_2} {c_3}
   \bigg] \, .
\eeqa
Further, using the balancing condition we have
\beqa
  h_i &=& 1+ \frac{2 \, r_0 \, \hat{h}_i  {{s_i}}^2}
    {r \, \hat{h}_1 \hat{h}_2 \hat{h}_3}
  \Bigg[
    \sinh^2{\sigma} - \sinh^2{\gamma_i} 
    + \frac{r_0}{2 r}
    \left(
      \cosh^2\sigma \sinh^2\gamma_i
      +\frac{\prod_{j=1}^3\sinh^2\gamma_j}{\sinh^2\gamma_i}
    \right)
  \Bigg]  .
\eeqa

\vfill

\newpage


\end{document}